\documentclass[12pt]{article}
\usepackage{stmaryrd}
\usepackage{bbm}
\usepackage{mathrsfs}
\usepackage{amssymb}
\usepackage{amsmath}
\usepackage{amsthm}
\usepackage{graphicx}
\usepackage{color}
\usepackage[square,numbers,sort&compress]{natbib}

\newtheorem{theorem}{Theorem}

\newtheorem{lemma}{Lemma}

\newcommand{\sdsum}{\rotatebox{90}{\rotatebox{90}{\rotatebox{90}{$\uplus$}}}}

\allowdisplaybreaks[4]
\begin{document}

\title{On contact symmetries of evolution equations
\thanks{Supported by the NSF for Distinguished Young
Scholars of China (Grant No. 10925104), the NSF of China (Grant Nos.
11101332, 11201371 and U1204104), the Foundation of Shaanxi
Educational Committee, China (Grant No. 11JK0482) and the NSF of Shaanxi Province, China (Grant No. 2012JQ1013).} }
\author{{ Qing Huang$^1$, Renat Zhdanov$^2$ and Changzheng Qu$^3$}\\
{\small 1. Department of Mathematics, Northwest University, Xi'an 710069, China}\\
{\small 2. BIO-key International, 55121 Eagan, MN, USA}\\
{\small 3. Department of Mathematics, Ningbo University, Ningbo 315211, China}}
\date{}
\maketitle

\begin{abstract}
  In this paper, we develop an algebraic approach to classifying contact symmetries of
the second-order nonlinear evolution equations. Up to contact isomorphisms, all inequivalent PDEs admitting
semi-simple algebras, solvable algebras of dimension $n\leq4$, and algebras having nontrivial
Levi factors, belonging to the class under consideration, and
corresponding contact symmetries they admitted are listed. 
\end{abstract}

\vskip 0.5in

\section{Introduction}
In this paper, we deals with the general evolution equations of the form
\begin{equation}\label{eq}
u_t=F(t,x,u,u_x,u_{xx}),
\end{equation}
in order to describe all possible functions $F$ such that partial differential equations (PDEs) \eqref{eq} admits nontrivial contact transformation group. By nontrivial contact symmetry group, we mean a contact transformation group which is at least one-parameter.
Hereafter, $u=u(t,x),u_t=\partial u/\partial t,u_x=\partial u/\partial x,u_{xx}=\partial^2 u/\partial x^2$, and $F$ is an
arbitrary sufficiently smooth real valued function of the indicated variables with $F_{u_{xx}}\neq0$.

The class of \eqref{eq} is a generalization of many important equations of mathematical physics. It contains, in particular, heat, Fisher, Newell-Whitehead, Burgers, Burgers-Huxley, Burgers-Fisher and Schr$\ddot{o}$dinger
equation. With specific $F$, Eq. \eqref{eq} has been used to model a wide variety of phenomena in physical, biological, economic science and many others \cite{los90, new69, yef04}.

Group classification has already become a universal and convenient tool for analysis of PDEs. There exist a wealth of excellent references on different aspects of group classification (see \cite{zhd99, zhd00, zhd01, zhd07} and references therein). The concept of transformation group is the core of Lie group approach. In the case when transformations only involve the dependent and independent variables, the corresponding transformation group is called point transformation group. The symmetry properties of point transformation of Eq. \eqref{eq} have been extensively investigated (see, e.g. Refs. \cite{zhd99, zhd00, zhd01, zhd07}).

If transformation laws for dependent or independent variables do contain gradients of dependent variables, then the corresponding group
is called contact. Roughly speaking, whereas there have been many publications devoted to the study of point symmetry structure of PDEs, the same is not true for contact symmetries. The notion of tangential transformation was first presented in Lie's doctoral thesis \cite{lie71}, in which he found a local contact transformation mapping straight lines into sphere in space. Application of contact symmetry to DEs was inaugurated by Lie himself, he proved that for ODEs of order $n$ ($n\geq3$), the contact symmetry algebra is finite-dimensional. He further classified linear PDE with two independent variables admitting contact symmetries \cite{lie81}. Since then, there have been some works on the theoretical foundation of contact transformations \cite{eis33, lie96} and its applications to geometry and dynamics. Contact symmetry algebras of the simple harmonic oscillator and scalar second-order ODEs are studied in \cite{mah91} and \cite{sch83} respectively. Contact transformations have also been applied to third-order ODEs to obtain hidden transformations \cite{abr95} and condition to admit an irreducible contact symmetries \cite{soh02}. What's more, they have been used in the linearization of third-order ODEs in \cite{ibr05}.

As for contact symmetries of PDEs, there are also some results. Contact symmetries of evolution equations are dealt with in \cite{sok87, sok88, mag93, mom99}. Contact transformations of Fokker-Planck \cite{rud99}, wave \cite{ibr00, mom01, mom03} and hyperbolic equations \cite{mor06} are investigated and classified. They have also been used in \cite{puc94} to obtain pseudo-invariant solutions of second-order PDEs. The relationship between potential and contact symmetries \cite{hua11,zhd09}, conservation law and contact symmetries \cite{kar00} are clarified. What's more, the concept of contact symmetry is generalized to discrete case in \cite{hyd98}.

Adapting the approach for Lie symmetries in \cite{zhd99} to contact symmetries, we develop an algebraic classification algorithm. This algorithm is implemented as three major steps. The first step is to compute the most general contact symmetry group of \eqref{eq} together with the classifying equation for $F$. In addition, we calculate the maximal contact equivalence group admitted by \eqref{eq}, namely, the contact transformation group preserving the class of PDEs \eqref{eq}.

The second step is essentially based on the explicit forms of commutation relations of low-dimensional abstract Lie algebras \cite{bar86, ndi75, pat76, tur88}. Using these, we describe all inequivalent realizations of contact symmetry algebras by basis infinitesimal operators admitted by \eqref{eq}.

At the third step, substituting the obtained realizations of low dimensional Lie algebras as symmetry algebras into the classifying equation, we derive the explicit forms of invariant equations. We also need to make sure that the corresponding symmetry algebras are maximal in Lie's sense.

The paper is organized as follows. In Section 2, we briefly present some necessary notions and results for contact transformations needed later. We also derive the classifying equation for the function $F$ and compute the maximal equivalence group of \eqref{eq}. Section 3 is devoted to a complete description of PDEs \eqref{eq} admitting semi-simple Lie algebras. In Section 4, 5 and 6, we carry out the classification of Eq. \eqref{eq} invariant with respect to two-, three- and four-dimensional solvable Lie algebras respectively. In Section 7, we obtain the equations invariant under algebras having nontrivial Levi factors. And Section 8 contains a summary and discussion.

\section{Contact transformation and classification equation}
Here we consider contact symmetries of evolution equations of the
form \eqref{eq}, we only summarize relevant aspects for the case of two independent variables
(time $t$ and one space variable $x$). The reader is referred to \cite{and79,ibr85,olv86}. It's common knowledge that the most general contact transformation
group admitted by \eqref{eq} is generated by the infinitesimal
operators
\begin{eqnarray}\label{Vg}
V_g&=&-g_{u_t}\partial_t-g_{u_x}\partial_x +
(g-u_tg_{u_t}-u_xg_{u_x})\partial_u + (g_x+u_xg_u)\partial_{u_x}\nonumber\\
&& + (g_t+u_tg_u)\partial_{u_t}.
\end{eqnarray}
where $g=g(t,x,u,u_t,u_x)$ is an arbitrary real-valued smooth
function, and $g$ is called generating, characteristic
function or contact Hamiltonian of $V_g$. Here and hereafter we shall identify the field
$V_g$ with its $g$. It's well known that contact transformation \eqref{Vg}
is equivalent to a smooth first-order Lie-B$\ddot{a}$cklund vector field (LBVF)
\begin{equation}\label{lb}
  g\partial_u+(D_xg)\partial_{u_x}+(D_tg)\partial_{u_t}+(D^2_xg)\partial_{u_{xx}}+\dots,
\end{equation}
where the symbols $D_t$ and $D_x$ denote the total differentiation operators with respect to the variables $t$ and $x$ respectively,
\begin{align*}
  D_t &= \partial_t+u_t\partial_u+u_{tt}\partial_{u_t}+u_{xt}\partial_{u_x}+\dots, \\[2mm]
  D_x &= \partial_x+u_x\partial_u+u_{tx}\partial_{u_t}+u_{xx}\partial_{u_x}+\dots,
\end{align*}
and $D^{j+1}_x=D_x(D^j_x)$. Furthermore, if function $g$ is of the form
\begin{equation*}
    g=\eta(t,x,u)-\tau(t,x,u)u_t-\xi(t,x,u)u_x,
\end{equation*}
i.e., $g_{u_tu_t}=g_{u_tu_x}=g_{u_xu_x}=0$, the contact transformation \eqref{Vg} and the LBVF \eqref{lb} are equivalent to the usual
Lie vector field
\begin{equation*}
  \tau(t,x,u)\partial_t+\xi(t,x,u)\partial_x+\eta(t,x,u)\partial_u.
\end{equation*}
As the form of a LBVF is uniquely determined by the function $g$, we often use the shorthand notation $g\partial_u$ instead of its full version \eqref{lb}. Eq. \eqref{eq} is invariant under the LBVF $g\partial_u$ or equivalently under the contact vector field \eqref{Vg}  iff the condition
\begin{equation*}
[gF_u+D_xgF_{u_x}+D^2_xgF_{u_{xx}}-D_tg]_{u_t=F}=0.
\end{equation*}
holds. The detailed description of the procedure for calculating LBVF of PDEs can be found in \cite{and79,ibr85,olv86}.
Substituting $u_t=F$ and its differential consequences into above invariance criterion yields an overdetermined system of linear
PDEs. Solving it, we arrive at the following assertion.
\begin{lemma}
The most general contact symmetry admitted by \eqref{eq} is generated by the
generating function
\begin{equation}\label{g}
    g(t,x,u,u_t,u_x)=\alpha(t)u_t+G(t,x,u,u_x),
\end{equation}
where $\alpha$ and $G$ are real-valued functions satisfying the
classifying equation
\begin{equation}\label{ce}
\begin{split}
    &-G_{u_x}F_uu_x+F_{u_{xx}}G_uu_{xx}+2F_{u_{xx}}G_{u_xx}u_{xx}+2F_{u_{xx}}G_{ux}u_x+F_{u_x}G_uu_x\\[2mm]
    &+F_{u_{xx}}G_{uu}u_x^2+F_{u_{xx}}G_{u_xu_x}u_{xx}^2+2F_{u_{xx}}u_xG_{u_xu}u_{xx}+F_uG+F_{u_{xx}}G_{xx}\\[2mm]
    &-G_{u_x}F_x+F_{u_x}G_x-\alpha F_t-FG_u-\alpha_tF-G_t=0.
\end{split}
\end{equation}
\end{lemma}

Now the contact symmetry group classification problem of Eq. \eqref{eq} reduces to constructing all possible solutions
of Eq. \eqref{ce}. The problem, however, is that \eqref{ce} is a underdetermined system of one
PDE for two unknown functions $\alpha$ and $G$. In addition, unknown function $F$ is to be determined as well.
To overcome this difficulty, we adopt the approach developed in \cite{zhd99,zhd07} and utilize
the classical results on classification of abstract low dimensional Lie algebras for constructing contact symmetries
of the form \eqref{g}. With explicit forms of symmetries in hand we can proceed to integrating
corresponding classifying equation and get the explicit forms of $F$.

We derive the maximal contact transformation group preserving Eq. \eqref{eq} directly and state the results as below.
\begin{lemma}
The most general contact equivalence group of Eq. \eqref{eq} reads as
\begin{equation}\label{tr}
 \tilde{t}=T(t),\quad \tilde{x}=X(t,x,u,u_x),\quad \tilde{u}=U(t,x,u,u_x)
\end{equation}
where the functions $X$ and $U$ satisfy the contact condition
\begin{equation}\label{cc}
X_{u_x}(u_xU_u+U_x)=U_{u_x}(u_xX_u+X_x).
\end{equation}
What's more, contact transformation \eqref{tr} and \eqref{cc} can convert the most general generating function
\begin{equation*}
\tilde{g}(\tilde{t},\tilde{x},\tilde{u},\tilde{u}_{\tilde{t}},\tilde{u}_{\tilde{x}})=
\tilde{\alpha}(\tilde{t})\tilde{u}_{\tilde{t}}+\tilde{G}(\tilde{t},\tilde{x},\tilde{u},\tilde{u}_{\tilde{x}})
\end{equation*}
of equation
\begin{equation*}
\tilde{u}_{\tilde{t}}=\tilde{F}(\tilde{t},\tilde{x},\tilde{u},\tilde{u}_{\tilde{x}},\tilde{u}_{\tilde{x}\tilde{x}})
\end{equation*}
to a new generating function
\begin{equation}\label{ng}
     g(t,x,u,u_t,u_x)=\frac{\tilde{\alpha}(T)}{\dot{T}}u_t+\frac{D(X)}{J}[ \frac{D(X)U_t-D(U)X_t}{D(X)\dot{T}}\tilde{\alpha}(T)
    +\tilde{G}(T,X,U,\tilde{u}_{\tilde{x}}) ]
\end{equation}
of the equation
\begin{equation*}
    u_t=\frac{D(X)}{J}[\dot{T}\tilde{F}(T,X,U,\tilde{u}_{\tilde{x}},\tilde{u}_{\tilde{x}\tilde{x}})+X_t\tilde{u}_{\tilde{x}}-U_t]
\end{equation*}
where
\begin{equation*}
    \tilde{u}_{\tilde{x}}=\frac{D(U)}{D(X)},\qquad \tilde{u}_{\tilde{x}\tilde{x}}=\frac{D(X)D^2(U)-D(U)D^2(X)}{[D(X)]^3}.
\end{equation*}
Here and elsewhere $D=\partial_x+\sum\limits_{i=0}^\infty u_{i+1}\partial_{u_i}$, $u_i=\partial^i u/\partial x^i$, $D^2=D\circ D$ and $J=D(X)U_u-D(U)X_u$.
\end{lemma}

Note that if symmetry, corresponding to the transformations \eqref{tr} which does not change the variable $t$, we call it evolution. The generator of
evolution is independent of $u_t$, ie. it has the form \eqref{g} with $\alpha(t)=0$. Evolution $g(x,u,u_x)$ for $n$-th order evolution
equations have been extensively studied in \cite{sok87, sok88}

Now we consider the cases $\tilde{\alpha}=0$ and
$\tilde{\alpha}\neq0$ separately.

Case 1. If $\tilde{\alpha}=0$, from \eqref{ng}, we get
\begin{equation*}
    g(t,x,u,u_t,u_x)=\frac{D(X)}{J}\tilde{G}(T,X,U,\frac{D(U)}{D(X)}).
\end{equation*}
It can be rectified to 1 by suitably choosing $X$ and $U$ satisfying
the contact condition \eqref{cc}.

For general $\tilde{G}$, we can not give out its corresponding $X$ and
$U$ explicitly. Here we take the case
$\tilde{G}=\tilde{u}_{\tilde{x}}$ as an example. At this time
\begin{equation*}
    g=\frac{D(X)}{J}\tilde{u}_{\tilde{x}}=\frac{D(U)}{J}=\frac1{\frac{D(X)}{D(U)}U_u-X_u},
\end{equation*}
By choosing the contact transformation \eqref{tr} and \eqref{cc} of the form
\begin{equation*}
    \tilde{t}=T(t),\quad \tilde{x}=-u+Y(t,x,U),\quad \tilde{u}=U(t,x,u,u_x),
\end{equation*}
where
\begin{equation*}
    Y_x=u_x,
\end{equation*}
we have
\begin{equation*}
    g=\frac1{\frac{(Y_x+Y_UU_x)+(-1+Y_UU_u)u_x}{U_x+U_uu_x}U_u-(-1+Y_UU_u)}=1.
\end{equation*}
For other $\tilde{G}$, the
analysis is similar.

Case 2. If $\tilde{\alpha}\neq0$, choosing $T$ satisfying the
equation $\dot{T}(t)=\tilde{\alpha}(T(t))$ yields
\begin{equation*}
    g=u_t+\frac{D(X)}{J}[
    \frac{D(X)U_t-D(U)X_t}{D(X)}
    +\tilde{G}(T,X,U,\frac{D(U)}{D(X)}) ],
\end{equation*}
By selecting $X$ and $U$ satisfying
\begin{equation*}
    \frac{D(X)U_t-D(U)X_t}{D(X)}
    +\tilde{G}(T,X,U,\frac{D(U)}{D(X)})=0,
\end{equation*}
and the contact condition \eqref{cc}, we obtain the generating
function
\begin{equation*}
    g=u_t
\end{equation*}

It is straight forward to verify that Lie-Backl$\ddot{u}$nd symmetry
operators $\partial_u$ and $u_t\partial_u$ are inequivalent.

\begin{lemma}\label{1-d}
Within the contact transformation \eqref{tr} and \eqref{cc}, the characteristic
function \eqref{g} is equivalent to one of the canonical
characteristic functions
\begin{equation*}
    1,\qquad u_t.
\end{equation*}
\end{lemma}

\begin{theorem}
There are two inequivalent equations \eqref{eq} invariant under
one-dimensional contact symmetry groups. The symmetries and their invariant equations are given by
\begin{align*}
    & A_1^1=\langle u_t \rangle :\quad u_t=F(x,u,u_x,u_{xx}),\\[3mm]
    & A_1^2=\langle 1 \rangle :\quad u_t=F(t,x,u_x,u_{xx}).
\end{align*}
\end{theorem}

To proceed any further, we need to define a new Lie bracket for functions $f(t,x,u,u_t,u_x)$ and $g(t,x,u,u_t,u_x)$. Let
\begin{equation*}
    f_{\ast}(g)=\partial_{u_t} f D_t g+\partial_u f g+\partial_{u_x}f D_x
    g,
\end{equation*}
and put
\begin{equation*}
    [f,g]=g_{\ast}(f)-f_{\ast}(g).
\end{equation*}
We can explicitly compute the right-hand side of
above formula and obtain
\begin{equation}\label{b}
\begin{split}
    &[f,g]=(\partial_{u_t}g\partial_{u}f-\partial_{u_t}f\partial_{u}g)u_t+(\partial_{u_x}g\partial_{u}f-\partial_{u_x}f\partial_{u}g)u_x\\[2mm]
    &\qquad\quad
    +\partial_{u_t}g\partial_{t}f-\partial_{u_t}f\partial_{t}g+\partial_{u_x}g\partial_{x}f-\partial_{u_x}f\partial_{x}g+f\partial_{u}g-g\partial_{u}f.
\end{split}
\end{equation}
From this formula it is clear that $[f,g]$ does not depend on
$u_{xx}$, $u_{tx}$ or $u_{tt}$, that is, all contact symmetries of
Eq.\eqref{eq} form a Lie algebra. We can also verify that
$[V_f,V_g]=V_{[f,g]}$ by direct computation, where $[V_f,V_g]$ is
the usual Lie bracket and $[f,g]$ is given by \eqref{b}.

The fundamental Levi-Malcev theorem says that, for an arbitrary finite-dimensional Lie algebra $L$
with the radical $N$ (the largest solvable ideal in $L$), there exists a semi-simple subalgebra $S$
such that
\begin{equation}\label{levi}
    L=S \sdsum N,
\end{equation}
where semi-simple subalgebra $S$ is called the Levi factor, and relation \eqref{levi} is called the Levi decomposition
of $L$. Consequently, Lie algebras fall into the following three categories: semi-simple algebras, solvable algebras, and semi-direct sums of solvable and semi-simple algebras. In the following sections, we study Eq. \eqref{eq} invariant under these three kinds of algebras step by step.

\section{Group classification of equations invariant under semi-simple Lie algebras}
In this section, we are devoted to constructing equations
whose invariance algebras are semi-simple. The lowest dimensional real
semi-simple Lie algebras are isomorphic to one of the following two
three-dimensional algebras \cite{bar86,ndi75}:
\begin{equation*}
\begin{array}{cl}
    {\frak{so}}(3): &[g_1,g_2]=g_3,\ [g_1,g_3]=-g_2,\ [g_2,g_3]=g_1;\\[2mm]
    {\frak{sl}}(2,\mathbf{R}): &[g_1,g_2]=g_1,\ [g_1,g_3]=2g_2,\
    [g_2,g_3]=g_3.
\end{array}
\end{equation*}
The following assertion holds.

\begin{theorem}\label{semisimple}
There are three inequivalent ${\frak{sl}}(2,\mathbf{R})$ algebras admitted by Eq.\ \eqref{eq}.
The maximal symmetry algebra and corresponding invariant equations read as
\begin{align*}
& {\frak{sl}}^1(2,\mathbf{R})=\langle 1,u,u^2-u_x^2 \rangle:\\[2mm]
& \qquad\qquad              u_t=u_xF(t,-x+\mathrm{arctanh}\frac{u_{xx}}{u_x}),\\[2mm]
& {\frak{sl}}^2(2,\mathbf{R})=\langle 1,u,u^2+u_x^2 \rangle:\\[2mm]
& \qquad\qquad              u_t=u_xF(t,x+\arctan{\frac{u_{xx}}{u_x}}),\\[2mm]
& {\frak{sl}}^3(2,\mathbf{R})=\langle u_t,-tu_t+xu_x,t^2u_t-2txu_x-u_x \rangle:\\[2mm]
& \qquad\qquad              u_t=-x^2u_x+\frac1{u_x}F(u,\frac{u_{xx}}{u_x^2}).
\end{align*}
And any ${\frak{so}}(3)$-invariant algebra is equivalent to
\begin{equation*}
{\frak{so}}^1(3)=\langle 1, \tan{x}\sin{u}-u_x\cos{u},\tan{x}\cos{u}+u_x\sin{u} \rangle,
\end{equation*}
it's invariant equation \eqref{eq} is
\begin{equation*}
    u_t=(\sec^2{x}+u_x^2)^\frac12 F(t,\frac{u_{xx}\cos{x}-(2+u_x^2\cos^2{x})u_x\sin{x}}{(1+u_x^2\cos^2{x})^\frac32}).
\end{equation*}
\end{theorem}

\proof
Consider first inequivalent realizations of the algebra ${\frak{sl}}(2,\mathbf{R})$. Taking
\eqref{g} as its basis $g_i\ (i=1,2,3)$, inserting them into the commutation relations
of ${\frak{sl}}(2,\mathbf{R})$, and solving equations obtained, we can get all possible realizations
of the algebra under study.

In view of Lemma \ref{1-d}, without loss of generality, we can assume that one of the basis contact Hamiltonian,
say $g_1$, can be reduced to one of the canonical forms $1$ and $u_t$.

Let $g_1=1$ and $g_2$, $g_3$ be of the form \eqref{g}. Inserting them into the first two commutation relations
of ${\frak{sl}}(2,\mathbf{R})$ yields
\begin{equation*}
    g_2=u+\phi(t,x,u_x),
\end{equation*}
here $\phi$ is an arbitrary real-valued smooth function. Before simplifying $g_2$, we seek those equivalence
transformations \eqref{tr} and \eqref{cc} which preserve the basis generating function $g_1=1$, namely,
\begin{equation*}
    \tilde{g_1}=1\rightarrow g_1=\frac{D(X)}{J}=1.
\end{equation*}
Hence,
\begin{equation}\label{e1}
    \frac{D(X)}{D(X)U_u-D(U)X_u}=1,
\end{equation}
and
\begin{equation}\label{e2}
X_{u_x}D(U)=U_{u_x}D(X).
\end{equation}
Eq.\ \eqref{e1} is equivalent to
\begin{equation}\label{e3}
    1=U_u-\frac{D(U)}{D(X)}X_u.
\end{equation}
Now we consider the two cases $X_{u_x}\neq0$ and $X_{u_x}=0$
separately.

Case 1. If $X_{u_x}\neq0$, we obtain $D(U)/D(X)=U_{u_x}/X_{u_x}$
from \eqref{e2}. Substituting it into \eqref{e3} yields the first
order linear PDE
\begin{equation*}
    X_{u_x}=X_{u_x}U_u-X_uU_{u_x}
\end{equation*}
for $U(t,x,u,u_x)$. Then, relation
\begin{equation*}
    \frac{{\rm d}t}0=\frac{{\rm d}x}0=\frac{{\rm d}u}{X_{u_x}}=\frac{{\rm
    d}u_x}{-X_u}=\frac{{\rm d}U}{X_{u_x}}
\end{equation*}
holds. Solving it, we obtain the following integral invariants
\begin{equation*}
    t,x,U-u,X.
\end{equation*}
Consequently, we can take $U-u=Y(t,x,X)$. Thus, we have
\begin{equation}\label{t1a}
    \tilde{x}=X(t,x,u,u_x),\qquad\tilde{u}=u+Y(t,x,X).
\end{equation}
Inserting above $X$ and $U$ into the contact condition \eqref{e2}
yields
\begin{equation*}
    X_{u_x}[(Y_x+Y_XX_x)+u_x(1+Y_XX_u)+u_{xx}Y_XX_{u_x}]=Y_XX_{u_x}[X_x+u_xX_u+u_{xx}X_{u_x}].
\end{equation*}
Namely, $X_{u_x}(u_x+Y_x)=0$ holds. Thus, we have
\begin{equation}\label{t1b}
Y_{x}=-u_x.
\end{equation}

Case 2. If $X_{u_x}=0$, we have $D(X)\neq0$ (otherwise $X=X(t)$).
Thus it follows from the contract condition \eqref{e2} that
$U_{u_x}=0$. Substituting
\begin{equation*}
    X=X(t,x,u),\qquad U=U(t,x,u),
\end{equation*}
into Eq. \eqref{e3} yields
\begin{equation*}
   1=U_u-\frac{U_x+u_xU_u}{X_x+u_xX_u}X_u,
\end{equation*}
namely,
\begin{equation*}
    X_x+u_xX_u=X_xU_u-X_uU_x.
\end{equation*}
Since functions $X$ and $U$ are independent of $u_x$, we have $X_u=0$.
Then we have $X=X(t,x)$ and above relation changes into
\begin{equation*}
    X_x=X_xU_u.
\end{equation*}
Consequently, we have $U_u=1$. Thus, point transformation
\begin{equation}\label{t2}
    \tilde{t}=T(t),\qquad \tilde{x}=X(t,x),\qquad \tilde{u}=u+Y(t,x),\qquad X_x\neq0
\end{equation}
is obtained.

Performing an inner automorphism transformation \eqref{t1a} and \eqref{t1b} preserving 1 to above $g_2$ yields
\begin{equation*}
    \tilde{g}_2=\tilde{u}+\phi(\tilde{t},\tilde{x},\tilde{u}_{\tilde{x}}) \rightarrow g_2=u+Y(t,x,X)+\phi(T,X,Y_X).
\end{equation*}
By choosing suitable $Y$ satisfying $Y(t,x,X)+\phi(T,X,Y_X)=0$ and \eqref{t1b}, we may suppose that
\begin{equation*}
  g_2=u.
\end{equation*}

Then it follows from the commutativity relations that $g_3=u^2+\psi(t,x)u_x^2$, where $\psi$ is an arbitrary
function. Applying transformations preserving 1 and $u$
\begin{equation*}
  \tilde{t}=T(t),\qquad \tilde{x}=X(t,x),\qquad \tilde{u}=u,\qquad X_x\neq0
\end{equation*}
to $\tilde{g_3}$ yields
\begin{equation*}
    \tilde{g_3}=\tilde{u}^2+\psi(\tilde{t},\tilde{x})\tilde{u}_{\tilde{x}}^2 \rightarrow g_3=u^2+\frac{\psi(T,X)}{X_x^2}u_x^2.
\end{equation*}
Now either $\psi=0$ or by a change of variable the function $\psi$ may be transformed to $1$ or $-1$. Thus,
three ${\frak{sl}}(2,\mathbf{R})$ algebras
\begin{equation*}
\langle 1,u,u^2 \rangle,\quad \langle 1,u,u^2-u_x^2 \rangle,\quad \langle 1,u,u^2+u_x^2 \rangle
\end{equation*}
are obtained.

Turn now to the case $g_1=u_t$, similar analysis yields two inequivalent realizations
of ${\frak{sl}}(2,\mathbf{R})$ algebras $\langle u_t,-tu_t+xu_x,t^2u_t-2txu_x-u_x \rangle$ and $\langle u_t,-tu_t,t^2u_t \rangle$.

The case of the algebra ${\frak{so}}(3)$ is treated in a similar way which gives one realization given
in the formulation of the theorem.

Inserting above obtained realizations into the classifying equation \eqref{ce}, we find the algebras
$\langle 1,u,u^2 \rangle$ and $\langle u_t,-tu_t,t^2u_t \rangle$ can not be admitted by Eq.\ \eqref{eq}
and also arrive at other invariant equations presented in the theorem. The assertion is proven. \qed

\begin{theorem}
The invariant equations listed in Theorem \ref{semisimple} exhaust the list of all
possible inequivalent PDEs \eqref{eq}, whose invariance algebras are
semi-simple.
\end{theorem}
\proof It is a common knowledge that there exist four basic types of
classical semi-simple Lie algebras, $A_{n-1},\ B_n,\ C_n,\ D_n$, and
five exceptional semi-simple Lie algebras, $G_2,\ F_4,\ E_6,\ E_7,\
E_8$ over the field of real numbers \cite{bar86, ndi75}.

$\bullet$ $A_{n-1}\ (n>1)$ has four real forms of the algebra
$\frak{sl}(n,\mathbf{C})$: $\frak{su}(n)$,
$\frak{sl}(n,\mathbf{R})$, $\frak{su}(p,q)\ (p+q=n,p\geq q)$,
$\frak{su}^{\ast}(2n)$.

$\bullet$ $B_n\ (n\geq1)$ contains two real forms of the algebra
$\frak{so}(2n+1,\mathbf{C})$: $\frak{so}(2n+1)$, $\frak{so}(p,q)\
(p+q=2n+1,p>q)$.

$\bullet$ $C_n\ (n\geq1)$ contains three real forms of the algebra
$\frak{sp}(n,\mathbf{C})$: $\frak{sp}(n)$,
$\frak{sp}(n,\mathbf{R})$, $\frak{sp}(p,q)\ (p+q=n,p\geq q)$.

$\bullet$ $D_n\ (n>1)$ has three real forms of the algebra
$\frak{so}(2n,\mathbf{C})$: $\frak{so}(2n)$, $\frak{so}(p,q)\
(p+q=2n,p\geq q)$, $\frak{so}^{\ast}(2n)$.

The lowest dimensional semi-simple Lie algebras admit the
following isomorphisms
\begin{equation*}
    \frak{sl}(2,\mathbf{R})\sim\frak{su}(1,1)\sim\frak{so}(2,1)
    \sim\frak{sp}(1,\mathbf{R}),\qquad
    \frak{so}(3)\sim\frak{su}(2)\sim\frak{sp}(1).
\end{equation*}
Hence, realizations of the algebra $\frak{sl}(2,\mathbf{R})$ and $\frak{so}(3)$
exhaust the set of all possible realizations of three-dimensional
semi-simple Lie algebras admitted by Eq.\ \eqref{eq}.

The next admissible dimension of a semi-simple Lie algebra
is six. There are four non-isomorphic semi-simple Lie algebras of the dimension six,
namely, $\frak{so}(4)$, $\frak{so}^{\ast}(4)$, $\frak{so}(2,2)$, and
$\frak{so}(3,1)$.

It's well known that $\frak{so}(4)\sim\frak{so}(3)\oplus\frak{so}(3)$. Consequently,
we can use the results of realizations of $\frak{so}(3)$ to classify $\frak{so}(4)$.
Now we need to construct all realizations of $\frak{so}(3)$ by \eqref{g}, which commute with
$\frak{so}^1(3)$. It's straightforward to verify that such realizations don't exist. Similarly,
for $\frak{so}^{\ast}(4)\sim\frak{so}(3)\oplus\frak{sl}(2,\mathbf{R})$, to classify $\frak{so}^{\ast}(4)$-invariant
equations \eqref{eq}, we have to construct all realizations of $\frak{sl}(2,\mathbf{R})$ by \eqref{g}, which commute
with $\frak{so}^1(3)$. The only realization of $\frak{sl}(2,\mathbf{R})$ obeying above constraints is equivalent to
$\langle u_t,-tu_t,t^2u_t \rangle$. The latter algebra cann't be invariance algebra of Eq.\ \eqref{eq}. Consequently,
there are no Eq.\ \eqref{eq} admitting $\frak{so}(4)$ and $\frak{so}^{\ast}(4)$ algebras.

As $\frak{so}(2,2)\sim\frak{sl}(2,\mathbf{R})\oplus\frak{sl}(2,\mathbf{R})$,
we can choose $\frak{so}(2,2)=\langle g_i,h_i|i=1,2,3\rangle$, where
$\langle g_1,g_2,g_3\rangle$ and $\langle
h_1,h_2,h_3\rangle$ are two $\frak{sl}(2,\mathbf{R})$ algebras with $[g_i,h_j]=0,\
(i,j=1,2,3)$. Taking $g_1,\ g_2,\ g_3$ to be the basis functions of
the realizations of $\frak{sl}(2,\mathbf{R})$ given in the
formulation of Theorem \ref{semisimple}, $h_1,\ h_2,\ h_3$ be of the
general form \eqref{g}, and analyzing the commutation relation of $\frak{so}(2,2)$,
we obtain the following three realizations of the algebra $\frak{so}(2,2)$,
\begin{align*}
& {\frak{sl}}^1(2,\mathbf{R})\oplus\langle u_t,-tu_t-\lambda u_x, t^2u_t+2\lambda tu_x \rangle,\\[2mm]
& {\frak{sl}}^2(2,\mathbf{R})\oplus\langle u_t,-tu_t-\lambda u_x, t^2u_t+2\lambda tu_x \rangle,\\[2mm]
& {\frak{sl}}^3(2,\mathbf{R})\oplus\langle 1,u,u^2 \rangle,
\end{align*}
with an arbitrary real constant $\lambda$. While these algebras can not be admitted by Eq.\ \eqref{eq}.

Algebra $\frak{so}(3,1)$ has the Cartan decomposition $\langle g_1,g_2,g_3\rangle\dot{+}\langle h_1,h_2,h_3\rangle$,
where $\langle g_1,g_2,g_3\rangle=\frak{so}(3)$, $[g_i,h_j]=\sum\limits_{k=1}^{3}\varepsilon_{ijk}h_k$, $[h_i,h_j]=-\sum\limits_{k=1}^{3}\varepsilon_{ijk}g_k$, $(i,j,k=1,2,3)$ and
\begin{equation*}
\varepsilon_{ijk}=\left\{\begin{array}{cl}
1   & \text{for even permutation of (i,j,k), i.e. 1,2,3; 2,3,1; 3,1,2}\\[3mm]
-1  & \text{for odd permutation of (i,j,k), i.e. 1,3,2; 3,2,1; 2,1,3}\\[3mm]
0   & \text{if there is a repeated index, i.e. 112, 322}.
\end{array}
\right.
\end{equation*}
Taking $\langle g_1,g_2,g_3\rangle=\frak{so}^1(3)$, simple calculation shows that the realization of
algebra $\frak{so}(3)$ cannot be extended to a realization of  $\frak{so}(3,1)$. Consequently, Eq.\ \eqref{eq}
does not admit six-dimensional semi-simple Lie algebras.

The same assertion holds true for eight-dimensional semi-simple Lie
algebras $\frak{sl}(3,\mathbf{R})$, $\frak{su}(3)$ and
$\frak{su}(2,1)$. Actually, it is also true for any semi-simple algebra of the dimension $n>3$.

As $\frak{su}^{\ast}(4)\sim\frak{so}(5,1)\supset\frak{so}(4)$, the
algebra $A_{n-1}\ (n>1)$ has no realizations by operators
\eqref{g} except for those given in Theorem \ref{semisimple}.
There are also no realizations of the algebra $D_n\ (n>1)$,
since the lowest dimensional algebras of this type $\frak{so}(4)$,
$\frak{so}(2,2)$, $\frak{so}^{\ast}(4)$ have no realizations within
the class of characteristic function \eqref{g}.

The similar reasoning yields that there are no new realizations of the
algebras $B_n\ (n>1)$ and $C_n\ (n\geq1)$ that can be symmetry
algebras of Eq.\eqref{eq}. Indeed, the algebra
$B_n$ contains $\frak{so}(4)$ and $\frak{so}(3,1)$ as subalgebras,
and what's more,
\begin{equation*}
 \frak{sp}(2)\sim\frak{so}(5)\supset\frak{so}(4),\
 \frak{sp}(2,\mathbf{R})\sim\frak{so}(3,2)\supset\frak{so}(3,1),\
 \frak{sp}(1,1)\sim\frak{so}(4,1)\supset\frak{so}(3,1).
\end{equation*}

It remains to consider the exceptional
semi-simple Lie algebras $G_2,F_4,E_6,E_7$ and $E_8$. Here we only
consider the first two algebras and others are handled in the same
way. The algebra $G_2$ contains a compact real form $g_2$ and one
noncompact real form $g_2'$, where $g_2\cap g_2'\sim
\frak{su}(2)\oplus\frak{su}(2)\sim\frak{so}(4)$. Since Eq.\eqref{eq}
cannot admit $\frak{so}(4)$ as an invariance algebra, $G_2$ has no
realization by operators \eqref{g}. A Lie algebra of type $F_4$ has a compact real form
$f_4$ and two noncompact real forms $f_4',\hat{f_4}$ with
$f_4\cap f_4'\sim
\frak{sp}(3)\oplus\frak{su}(2)\supset\frak{so}(3)$, $f_4\cap
\hat{f_4}\sim \frak{sp}(9)$. Thus Eq.\eqref{eq} cannot admit a
invariance algebra of type $F_4$. The same assertion
holds true for the remaining exceptional semi-simple Lie algebras
$E_6,E_7$ and $E_8$.

The theorem is proved.\qed

\section{Group classification of equations invariant under two-dimensional Lie algebras}

There are two non-isomorphic two-dimensional Lie algebras,
\begin{equation*}
    A_{2.1}:\ \ [g_1,g_2]=0,\qquad A_{2.2}:\ \ [g_1,g_2]=g_2.
\end{equation*}

As both $A_{2.1}$ and $A_{2.2}$ contain the algebra $A_1$, we can
assume that the basis function of the latter is reduced to a
canonical form. We consider in detail the case of $A_{2.1}$, while for the case of
algebra $A_{2.2}$ we present the final results only.

Let $g_1=1$ and $g_2$ be of the
most generic form \eqref{g},
\begin{equation*}
    g_2=\alpha(t)u_t+G(t,x,u,u_x).
\end{equation*}
Then the commutation relation implies that $G_u=0$, we can put
$g_2=\alpha(t)u_t+G(t,x,u_x)$.

Performing an inner automorphism transformation \eqref{t1a} and \eqref{t1b} preserving 1 to above $g_2$ yields
\begin{equation*}
    \tilde{g_2}= \tilde{\alpha}(\tilde{t})\tilde{u}_{\tilde{t}}+\tilde{G}(\tilde{t},\tilde{x},\tilde{u}_{\tilde{x}})\rightarrow
    g_2=\frac{\tilde{\alpha}(T)}{\dot{T}}u_t+\frac{D(X)U_t-D(U)X_t}{D(X)\dot{T}}\tilde{\alpha}(T)
    +\tilde{G}(T,X,\frac{D(U)}{D(X)})
\end{equation*}

Now if $\tilde{\alpha}=0$, we can arrive at
\begin{equation}\label{1}
    \tilde{g_2}=\tilde{G}(\tilde{t},\tilde{x},\tilde{u}_{\tilde{x}})\rightarrow
    g_2=\tilde{G}(T,X,\frac{D(U)}{D(X)}),
\end{equation}
We consider two cases separately.

Case 1. If
$\tilde{G}_{\tilde{x}}=\tilde{G}_{\tilde{u}_{\tilde{x}}}=0$, then
$\tilde{G}=\tilde{G}(\tilde{t})$. We reduce $\tilde{G}(T)$ by
selecting suitable point transformation \eqref{t2}.

Now if $\tilde{G}_{\tilde{t}}\neq0$,
$\tilde{G}=\tilde{G}(\tilde{t})$, by choosing the transformation
\eqref{t2} satisfying $t=\tilde{G}(\tilde{t})$ yields $g_2=t$.

Else if $\tilde{G}_{\tilde{t}}=0$, $g_2$ is a constant, and it
can't expand a two-dimension Lie algebra together with 1.

Case 2. If $\tilde{G}_{\tilde{x}}\neq0$ or
$\tilde{G}_{\tilde{u}_{\tilde{x}}}\neq0$, we can reduce $\tilde{g_2}$ to $g_2=u_x$ by suitable contact transformation \eqref{t1a} and \eqref{t1b}. Now we
establish the compatibility condition of the system
\begin{equation}\label{3}
\left\{\begin{array}{l}
     \tilde{G}(T,X,Y_X)=u_x,\\[3mm]
     Y_{x}=-u_x.
\end{array}\right.
\end{equation}
Note that here $\tilde{u}_{\tilde{x}}=Y_X$ according to the contact transformation \eqref{t1a} and
\eqref{t1b}.

System \eqref{3} is compatible iff
\begin{equation}\label{com}
    Y_{xX}=Y_{Xx}.
\end{equation}
In view of $X=X(t,x,u,u_x)$ where $X_{u_x}\neq0$ and applying the
implicit function theorem, we obtain that
\begin{equation*}
    u_x=u_x(t,x,u,X),
\end{equation*}
where
\begin{equation*}
    \frac{\partial u_x}{\partial x}=-\frac{X_x}{X_{u_x}},\qquad
    \frac{\partial u_x}{\partial X}=\frac1{X_{u_x}}.
\end{equation*}
Thus,
\begin{equation}\label{4}
    Y_{xX}=\frac{\partial Y_x}{\partial X}=-\frac{\partial u_x}{\partial
    X}=-\frac1{X_{u_x}}.
\end{equation}
\indent And now we derive the expression of $Y_{Xx}$.
Differentiating the first equation of \eqref{3} with respect to $x$
yields
\begin{equation*}
    \tilde{G}_{Y_X}Y_{Xx}=\frac{\partial u_x}{\partial
    x}=-\frac{X_x}{X_{u_x}}.
\end{equation*}
Consequently, we get
\begin{equation}\label{5}
    Y_{Xx}=-\frac1{\tilde{G}_{Y_X}}\frac{X_x}{X_{u_x}}.
\end{equation}
Combining \eqref{com}, \eqref{4} and \eqref{5} together, we can get the
assertion that system \eqref{3} is compatible iff
\begin{equation}\label{XG}
    X_x=\tilde{G}_{Y_X}=\tilde{G}_{\tilde{u}_{\tilde{x}}}.
\end{equation}

So if we choose $X$ and $Y$ in \eqref{t1a} and \eqref{t1b}
as solutions of compatible system of PDEs \eqref{3} and \eqref{XG},
$g_2=u_x$ can be obtained from \eqref{1}. Following two examples illustrate this assertion.

{\it Example 1.} Consider the characteristic function
\begin{equation*}
\tilde{g_2}=\tilde{x}.
\end{equation*}
In this case, system of Eqs.\ \eqref{3} and \eqref{XG} turns into
\begin{equation*}
    X=u_x,\qquad Y_x=-u_x,\qquad X_x=0.
\end{equation*}
Without loss of generality, we obtain the Legendre contact
transformation
\begin{equation*}
    \tilde{x}=u_x,\qquad \tilde{u}=u-xu_x.
\end{equation*}
Applying this transformation to $\tilde{g_2}=\tilde{x}$ in \eqref{1}, the new characteristic function
$g_2=u_x$ is obtained.

{\it Example 2.} Consider the characteristic function
\begin{equation*}
     \tilde{g_2}=\tilde{u}_{\tilde{x}}^2.
\end{equation*}
To reduce it to $g_2=u_x$, we have to find $X$ and $Y$ in \eqref{t1a}
and \eqref{t1b} satisfying the system
\begin{equation*}
 Y_X=u_x^\frac12,\qquad Y_x=-u_x,\qquad X_x=\tilde{G}_{Y_X}.
\end{equation*}
Rewriting $\tilde{G}$ as $\tilde{G}=Y_X^2$, we get
$X_x=2Y_X=2u_x^\frac12$. Without loss of generality, we can choose
\begin{equation*}
    X=2xu_x^\frac12.
\end{equation*}
Thus,
\begin{equation*}
    Y_X=\frac X{2x},\qquad Y_x=-\frac{X^2}{4x^2}.
\end{equation*}
Solving above equations, we obtain $Y=X^2/(4x)$. So, by choosing the contact transformation
\begin{equation*}
    \tilde{x}=2xu_x^\frac12,\qquad \tilde{u}=u+xu_x,
\end{equation*}
we can reduce $\tilde{g_2}=\tilde{u}_{\tilde{x}}^2$ to $g_2=u_x$.

For other $\tilde{g_2}$, the choice of $X$ and $U$ is similar to above two examples.

From above two cases, we have two two-dimensional Abelian algebras $\langle 1,u_x \rangle$ and $\langle 1,t \rangle$.
It can be easy shown that the algebra $\langle 1,t \rangle $ cann't
be admitted by Eq.\ \eqref{eq}.

And else if $\tilde{\alpha}\neq0$, choosing $T$ as solution
of equation $\dot{T}(t)=\tilde{\alpha}(T(t))$ in the transformation
\eqref{t1a} and \eqref{t1b} yields
\begin{equation}\label{2}
    \tilde{g_2}=\tilde{\alpha}(\tilde{t})\tilde{u}_{\tilde{t}}+\tilde{G}(\tilde{t},\tilde{x},\tilde{u}_{\tilde{x}})
    \rightarrow
    g_2=u_t+Y_t+\tilde{G}(T,X,Y_X).
\end{equation}
Now we prove that above generator can be reduced to $u_t$ by
suitable choosing $X$ and $Y$. To this end, we establish the
compatibility condition of the system
\begin{equation}\label{6}
\left\{\begin{array}{l}
     \tilde{G}(T,X,Y_X)+Y_t(t,x,X)=0,\\[3mm]
     u_x+Y_{x}(t,x,X)=0.
\end{array}\right.
\end{equation}

System \eqref{6} is compatible iff
\begin{equation}\label{6.1}
    Y_{tx}=Y_{xt},\quad Y_{Xx}=Y_{xX}.
\end{equation}
Differentiating the first equation of \eqref{6} with respect to $x$ yields
\begin{equation}\label{6.2}
    \tilde{G}_{Y_X}Y_{Xx}+Y_{tx}=0.
\end{equation}
In view of $X=X(t,x,u,u_x)$ where $X_{u_x}\neq0$ and applying the
implicit function theorem, we obtain that
\begin{equation*}
    u_x=u_x(t,x,u,X),
\end{equation*}
where
\begin{equation*}
    \frac{\partial u_x}{\partial t}=-\frac{X_t}{X_{u_x}},\qquad
    \frac{\partial u_x}{\partial X}=\frac1{X_{u_x}}.
\end{equation*}
Consequently,
\begin{equation}\label{6.3}
    Y_{xt}=\frac{X_t}{X_{u_x}},\quad Y_{xX}=-\frac1{X_{u_x}}.
\end{equation}
Combining \eqref{6.1}, \eqref{6.2} and \eqref{6.3} together yields
\begin{equation}\label{6.4}
    X_t=\tilde{G}_{Y_X}=\tilde{G}_{\tilde{u}_{\tilde{x}}}.
\end{equation}

So if we choose $X$ and $Y$ as solutions of the compatible system of
PDEs \eqref{6} and \eqref{6.4}, $\tilde{g_2}$ can be
reduced to $g_2=u_t$ according to \eqref{2}. Thus, algebra $\langle 1,u_t \rangle $ is
obtained.

We continue to the case $g_1=u_t$ and $g_2$ be the most generic form \eqref{g}.
By similar procedure, algebra $\langle u_t,1 \rangle $ is
constructed, while it is obtained before.

Now we have considered in detail the case of $A_{2.1}$, while for
algebra $A_{2.2}$, we present the final results only.
\begin{theorem}\label{2D}
There exist two Abelian and three non-Abelian two-dimensional
symmetry algebras admitted by \eqref{eq}. These algebras and the
corresponding invariant equations are given below,
\begin{align*}
    & A_{2.1}^1=\langle 1,u_x \rangle :\quad u_t=F(t,u_x,u_{xx}),\\[2mm]
    & A_{2.1}^2=\langle 1,u_t \rangle :\quad u_t=F(x,u_x,u_{xx}),\\[2mm]
    & A_{2.2}^1=\langle -u,1 \rangle :\quad u_t=u_xF(t,x,\frac{u_{xx}}{u_x}),\\[2mm]
    & A_{2.2}^2=\langle u_t-u,1 \rangle :\quad u_t=\mathrm{e}^{t}F(x,\mathrm{e}^{-t}u_x,\mathrm{e}^{-t}u_{xx}),\\[2mm]
    & A_{2.2}^3=\langle tu_t+1,u_t \rangle :\quad
    u_t=\mathrm{e}^{u}F(x,u_x,u_{xx}).
\end{align*}
\end{theorem}

\section{Equations admitting three-dimensional solvable Lie
algebras}

We consider the cases of decomposable and non-decomposable Lie
algebras separately.

\subsection{Three-dimensional decomposable algebras}
There exist two non-isomorphic three-dimensional decomposable Lie
algebras, $A_{3.1}$ and $A_{3.2}$,
\begin{equation*}
\begin{array}{l}
    A_{3.1}:\ \ [g_i,g_j]=0\ (i,j=1,2,3)\quad
    (A_{3.1}=A_1\oplus A_1\oplus A_1),\\[2mm]
    A_{3.2}:\ \ [g_1,g_2]=g_2,\ [g_1,g_3]=[g_2,g_3]=0,\quad
    (A_{3.2}=A_{2.2}\oplus A_1).
\end{array}
\end{equation*}

It is well-known that any three-dimensional solvable Lie
algebra contains two-dimensional solvable algebra. So that to
describe all possible realizations of three-dimensional solvable
algebras admitted by Eq. \eqref{eq}, it suffices to consider all
possible extensions of two dimensional algebras listed in Theorem
\ref{2D} by characteristic function $g_3$ of the form \eqref{g}. Then for
each of the so obtained realization we simplify $g_3$ using
equivalence transformations which preserve $g_1$ and
$g_2$. Having obtained explicit forms of symmetries, we can proceed to insert them into the classifying equation and integrating
corresponding equations. With these three steps performed, we obtain the following list
of invariant equations \eqref{eq}.

$A_{3.1}-$ invariant equations
\begin{align*}
    &A_{3.1}^1=\langle 1,u_x,u_t \rangle:\quad u_t=F(u_x,u_{xx}),\\[2mm]
    &A_{3.1}^2=\langle 1,u_x,h(t,u_x) \rangle,\ h_th_{u_xu_x}\neq0:\quad u_t=-\frac{h_t}{h_{u_xu_x}u_{xx}}+F(t,u_x).
\end{align*}

$A_{3.2}-$ invariant equations
\begin{align*}
    &A_{3.2}^1=\langle -u,1,u_x \rangle:\quad u_t=u_xF(t,\frac{u_{xx}}{u_x}),\\[2mm]
    &A_{3.2}^2=\langle -u,1,u_t \rangle:\quad u_t=u_xF(x,\frac{u_{xx}}{u_x}),\\[2mm]
    &A_{3.2}^3=\langle u_t-u,1,u_x \rangle :\quad u_t=\mathrm{e}^{t}F(\mathrm{e}^{-t}u_x,\mathrm{e}^{-t}u_{xx}),\\[2mm]
    &A_{3.2}^4=\langle tu_t+1,u_t,u_x \rangle :\quad
       u_t=\mathrm{e}^{u}F(u_x,u_{xx}).
\end{align*}

\subsection{Three-dimensional non-decomposable algebras}

There are seven non-isomorphic non-decomposable three-dimensional
real solvable Lie algebras. The list of these algebras is exhausted
by one nilpotent Lie algebra (only nonzero commutation relations
are given),
\begin{equation*}
    A_{3.3}:\ \ [g_2,g_3]=g_1,
\end{equation*}
and six solvable Lie algebras,
\begin{align*}
&A_{3.4}:\ \ [g_1,g_3]=g_1,\quad [g_2,g_3]=g_1+g_2,\\[2mm]
&A_{3.5}:\ \ [g_1,g_3]=g_1,\quad [g_2,g_3]=g_2,\\[2mm]
&A_{3.6}:\ \ [g_1,g_3]=g_1,\quad [g_2,g_3]=-g_2,\\[2mm]
&A_{3.7}:\ \ [g_1,g_3]=g_1,\quad [g_2,g_3]=qg_2,\ (0<|q|<1),\\[2mm]
&A_{3.8}:\ \ [g_1,g_3]=-g_2,\quad [g_2,g_3]=g_1,\\[2mm]
&A_{3.9}:\ \ [g_1,g_3]=qg_1-g_2,\quad [g_2,g_3]=g_1+qg_2,\ (q>0).
\end{align*}
All these algebras contain a two-dimensional Abelian ideal as a
subalgebra. Thus we can use our classification of $A_{2.1}-$
invariant equations to construct Eq.\eqref{eq}, which admit
non-decomposable three-dimensional solvable Lie algebras. We skip
intermediate calculations and present the final list of
invariant equations and the corresponding symmetry algebras.

$A_{3.3}-$ invariant equations
\begin{align*}
    & A_{3.3}^1=\langle 1,u_x,u_t-x \rangle :\quad u_t=F(u_x-t,u_{xx}),\\[2mm]
    & A_{3.3}^2=\langle 1,u_x,-x \rangle :\quad u_t=F(t,u_{xx}),\\[2mm]
    & A_{3.3}^3=\langle 1,u_t,u_x-t \rangle :\quad
    u_t=x+F(u_x,u_{xx}).
\end{align*}

$A_{3.4}-$ invariant equations
\begin{align*}
    & A_{3.4}^1=\langle 1,u_x,u-(1+u_x)x \rangle :\quad u_t=\mathrm{e}^{-u_x}F(t,\mathrm{e}^{-u_x}u_{xx}),\\[2mm]
    & A_{3.4}^2=\langle 1,u_x,u_t+u-(1+u_x)x \rangle :\quad u_t=\mathrm{e}^{-t}F(u_x-t,\mathrm{e}^{-t}u_{xx}),\\[2mm]
    & A_{3.4}^3=\langle 1,u_t,-tu_t+u-t \rangle :\quad u_t=-\ln{u_x}+F(x,\frac{u_{xx}}{u_x}).
\end{align*}

$A_{3.5}-$ invariant equations
\begin{align*}
    & A_{3.5}^1=\langle 1,u_x,u-xu_x \rangle :\quad u_t=\frac{F(t,u_x)}{u_{xx}},\\[2mm]
    & A_{3.5}^2=\langle 1,u_x,u_t+u-xu_x \rangle :\quad u_t=\mathrm{e}^{-t}F(u_x,\mathrm{e}^{-t}u_{xx}),\\[2mm]
    & A_{3.5}^3=\langle 1,u_t,-tu_t+u \rangle :\quad u_t=F(x,\frac{u_{xx}}{u_x}).
\end{align*}

$A_{3.6}-$ invariant equations
\begin{align*}
    & A_{3.6}^1=\langle 1,u_x,u+xu_x \rangle :\quad u_t=u_{x}^\frac12F(t,\frac{u_{xx}}{u_x^\frac32}),\\[2mm]
    & A_{3.6}^2=\langle 1,u_x,u_t+u+xu_x \rangle :\quad u_t=\mathrm{e}^{-t}F(\mathrm{e}^{2t}u_x,\mathrm{e}^{3t}u_{xx}),\\[2mm]
    & A_{3.6}^3=\langle 1,u_t,tu_t+u \rangle :\quad u_t=u_x^2F(x,\frac{u_{xx}}{u_x}).
\end{align*}

$A_{3.7}-$ invariant equations
\begin{align*}
    & A_{3.7}^1=\langle 1,u_x,u-qxu_x \rangle :\quad u_t=u_{x}^\frac1{1-q}F(t,u_x^\frac{1-2q}{q-1}u_{xx}),\\[2mm]
    & A_{3.7}^2=\langle 1,u_x,u_t+u-qxu_x \rangle :\quad u_t=\mathrm{e}^{-t}F(\mathrm{e}^{(1-q)t}u_x,\mathrm{e}^{(1-2q)t}u_{xx}),\\[2mm]
    & A_{3.7}^3=\langle 1,u_t,-qtu_t+u \rangle :\quad u_t=u_x^{1-q}F(x,\frac{u_{xx}}{u_x}),\\[2mm]
    & A_{3.7}^4=\langle u_t,1,-tu_t+qu \rangle :\quad
    u_t=u_x^\frac{q-1}qF(x,\frac{u_{xx}}{u_x}).
\end{align*}

$A_{3.8}-$ invariant equations
\begin{align*}
    & A_{3.8}^1=\langle 1,u_x,-x-uu_x \rangle :\quad u_t=(1+u_x^2)^\frac12F(t,\frac{u_{xx}}{(1+u_x^2)^\frac32}),\\[2mm]
    & A_{3.8}^2=\langle 1,u_x,u_t-x-uu_x \rangle :\quad
    u_t=(1+u_x^2)^\frac12F(t-\arctan{u_x},\frac{u_{xx}}{(1+u_x^2)^\frac32}).
\end{align*}

$A_{3.9}-$ invariant equations
\begin{align*}
    & A_{3.9}^1=\langle 1,u_x,(q-u_x)u-(1+qu_x)x \rangle :\\[2mm]
    &\qquad\qquad u_t=(1+u_x^2)^\frac12\mathrm{e}^{-q\arctan{u_x}}F(t,\frac{\mathrm{e}^{-q\arctan{u_x}}u_{xx}}{(1+u_x^2)^\frac32}),\\[2mm]
    & A_{3.9}^2=\langle 1,u_x,u_t+(q-u_x)u-(qu_x+1)x \rangle :\\[2mm]
    &\qquad\qquad u_t=(1+u_x^2)^\frac12\mathrm{e}^{-qt}F(\omega_1,\omega_3),\\[2mm]
    &\qquad\qquad\qquad \omega_1=t-\arctan{u_x},\\[2mm]
    &\qquad\qquad\qquad \omega_2=\frac12\sin{2t}\sin{2\omega_1}+2\cos^2t\cos^2{\omega_1}+\sin^2t-\cos^2{\omega_1},\\[2mm]
    &\qquad\qquad\qquad \omega_3=\mathrm{e}^{-qt}\omega_2^\frac32u_{xx}.
\end{align*}

\section{Classification of equations invariant under
four-dimensional solvable Lie algebras}

Now we perform group classification of Eq. \eqref{eq} admitting
four-dimensional solvable Lie algebras. To this end we use the
realizations of three-dimensional solvable algebras obtained in the
previous section and the fact that any four-dimensional solvable Lie
algebra contains a three-dimensional solvable algebra as a subalgebra.
We consider two cases of decomposable and non-decomposable Lie
algebras separately.

\subsection{Four-dimensional decomposable
algebras}

The list of non-isomorphic four-dimensional decomposable algebras
contains following ten algebras:
\begin{align*}
& A_{2.2}\oplus A_{2.2}=2A_{2.2},\\[2mm]
& A_{3.1}\oplus A_1=4A_1,\\[2mm]
& A_{3.2}\oplus A_1=A_{2.2}\oplus 2A_1,\\[2mm]
& A_{3.i}\oplus A_1\ (i=3,4,\cdots,9).
\end{align*}
The complete list of Eq. \eqref{eq} invariant with respect to above algebras is given below.

$2A_{2.2}-$ invariant equations
\begin{align*}
    & 2A_{2.2}^1=\langle -u,1,u_x,\mathrm{e}^{-x}u_x \rangle :\quad u_t=\frac{u_x^2}{u_{xx}-u_x}F(t),\\[2mm]
    & 2A_{2.2}^2=\langle -u,1,xu_x,u_x \rangle :\quad u_t=\frac{u_x^2}{u_{xx}}F(t),\\[2mm]
    & 2A_{2.2}^3=\langle -u,1,u_t+xu_x,u_x \rangle :\quad u_t=\mathrm{e}^tu_xF(\frac{\mathrm{e}^tu_{xx}}{u_x}),\\[2mm]
    & 2A_{2.2}^4=\langle -u,1,u_t,\mathrm{e}^{-t}u_x \rangle :\quad u_t=xu_x+u_xF(\frac{u_{xx}}{u_x}),\\[2mm]
    & 2A_{2.2}^5=\langle -u,1,u_t,\mathrm{e}^{-t}(u_t+u_x) \rangle :\quad u_t=-u_x+\mathrm{e}^xu_xF(\frac{u_{xx}}{u_x}),\\[2mm]
    & 2A_{2.2}^6=\langle -u,1,tu_t+u_x,u_t \rangle :\quad u_t=\mathrm{e}^{-x}u_xF(\frac{u_{xx}}{u_x}),\\[2mm]
    & 2A_{2.2}^7=\langle u_t-u,1,u_x,\mathrm{e}^{t-x} \rangle :\quad u_t=-u_x+\mathrm{e}^tF(\mathrm{e}^{-t}(u_x+u_{xx})),\\[2mm]
    & 2A_{2.2}^8=\langle u_t-u,1,\lambda u_t+xu_x,u_x \rangle,\ \lambda\neq-1 :\quad
         u_t=\mathrm{e}^{\frac{t}{1+\lambda}}u_x^{\frac{\lambda}{1+\lambda}}F(\mathrm{e}^{\frac{t}{1+\lambda }}u_x^{-\frac{2+\lambda}{1+\lambda}}u_{xx}),\\[2mm]
    & 2A_{2.2}^9=\langle u_t-u,1,-u_t+xu_x,u_x \rangle :\quad u_t=\frac{\mathrm{e}^{2t}}{u_{xx}}F(\mathrm{e}^{-t}u_x),\\[2mm]
    & 2A_{2.2}^{10}=\langle tu_t+1,u_t,u_x,\mathrm{e}^{-x} \rangle :\quad u_t=\mathrm{e}^{u+u_x}F(u_{xx}+u_x),\\[2mm]
    & 2A_{2.2}^{11}=\langle tu_t+1,u_t,xu_x,u_x \rangle :\quad
    u_t=\mathrm{e}^uF(\frac{u_{xx}}{u_x^2}).
\end{align*}

$A_{3.1}\oplus A_1-$ invariant equations
\begin{align*}
    &A_{3.1}^2\oplus A_1=\langle 1,u_x,g(t,u_x),h(t,u_x) \rangle,\ h_tg_{u_xu_x}=g_th_{u_xu_x}\neq0,g\neq C_1+C_2u_x+
     C_3h:\\[2mm]
    &\qquad\qquad u_t=-\frac{h_t}{h_{u_xu_x}u_{xx}}+F(t,u_x).
\end{align*}

$A_{3.2}\oplus A_1-$ invariant equations
\begin{align*}
     &A_{3.2}^1\oplus A_1=\langle -u,1,u_x,u_t \rangle:\quad u_t=u_xF(\frac{u_{xx}}{u_x}),\\[2mm]
     &A_{3.2}^3\oplus A_1=\langle u_t-u,1,u_x,\mathrm{e}^th(\mathrm{e}^{-t}u_x) \rangle,\ h''\neq0 :\\[2mm]
     &\qquad\qquad u_t=\frac{\mathrm{e}^t(u_xh'-\mathrm{e}^th)}{h''u_{xx}}+\mathrm{e}^{t}F(\mathrm{e}^{-t}u_x),\\[2mm]
     &A_{3.2}^4\oplus A_1=\langle tu_t+1,u_t,u_x,h(u_x) \rangle,\
          h''\neq0 :\quad u_t=\mathrm{e}^{u+\frac{h-u_xh'}{h''u_{xx}}}F(u_x).
     \end{align*}

$A_{3.3}\oplus A_1-$ invariant equations
\begin{align*}
    & A_{3.3}^1\oplus A_1=\langle 1,u_x,u_t-x,h(u_x-t) \rangle,\ h''\neq0 :\quad u_t=\frac{h'}{h''u_{xx}}+F(u_x-t),\\[2mm]
    & A_{3.3}^2\oplus A_1=\langle 1,u_x,-x,u_t \rangle :\quad u_t=F(u_{xx}),\\[2mm]
    & A_{3.3}^3\oplus A_1=\langle 1,u_t,u_x-t,h(u_x) \rangle,\ h''\neq0 :\quad u_t=-\frac{h'}{h''u_{xx}}+x+F(u_x),\\[2mm]
    & A_{3.3}^3\oplus A_1=\langle 1,u_t,u_x-t,u_t-x \rangle :\quad
    u_t=x+F(u_{xx}).
\end{align*}

$A_{3.4}\oplus A_1-$ invariant equations
\begin{align*}
    & A_{3.4}^1\oplus A_1=\langle 1,u_x,u-(1+u_x)x,\mathrm{e}^{-u_x}h(t) \rangle,\ h'\neq0 :\quad u_t=-\frac{h'}{u_{xx}h}+\mathrm{e}^{-u_x}F(t),\\[2mm]
    & A_{3.4}^1\oplus A_1=\langle 1,u_x,u-(1+u_x)x,u_t \rangle,\ h'\neq0 :\quad u_t=\mathrm{e}^{-u_x}F(\mathrm{e}^{-u_x}u_{xx}),\\[2mm]
    & A_{3.4}^2\oplus A_1=\langle 1,u_x,u_t+u-(1+u_x)x,\mathrm{e}^{-t}h(u_x-t) \rangle,\ (h+h')h''\neq0 :\\[2mm]
    &\qquad\qquad u_t=\frac{h+h'}{u_{xx}h''}+\mathrm{e}^{-t}F(u_x-t),\\[2mm]
    & A_{3.4}^3\oplus A_1=\langle 1,u_t,-tu_t+u-t,u_x \rangle :\quad u_t=-\ln{u_x}+F(\frac{u_{xx}}{u_x}).
\end{align*}

$A_{3.5}\oplus A_1-$ invariant equations
\begin{align*}
    & A_{3.5}^1\oplus A_1=\langle 1,u_x,u-xu_x,u_t \rangle :\quad u_t=\frac{F(u_x)}{u_{xx}},\\[2mm]
    & A_{3.5}^2\oplus A_1=\langle 1,u_x,u_t+u-xu_x,\mathrm{e}^{-t}h(u_x) \rangle,\ h''\neq0 :\quad u_t=\frac{h}{h''u_{xx}}+\mathrm{e}^{-t}F(u_x),\\[2mm]
    & A_{3.5}^3\oplus A_1=\langle 1,u_t,-tu_t+u,u_x \rangle :\quad u_t=F(\frac{u_{xx}}{u_x}).
\end{align*}

$A_{3.6}\oplus A_1-$ invariant equations
\begin{align*}
    & A_{3.6}^1\oplus A_1=\langle 1,u_x,u+xu_x,u_x^\frac12h(t) \rangle, \ h'\neq0 :\quad u_t=\frac{4h'u_x^2}{hu_{xx}}+u_{x}^\frac12F(t),\\[2mm]
    & A_{3.6}^1\oplus A_1=\langle 1,u_x,u+xu_x,u_t \rangle :\quad u_t=u_{x}^\frac12F(\frac{u_{xx}}{u_x^\frac32}),\\[2mm]
    & A_{3.6}^2\oplus A_1=\langle 1,u_x,u_t+u+xu_x,\mathrm{e}^{-t}h(\mathrm{e}^{2t}u_x) \rangle,\ h''\neq0,h(y)\neq y^\frac12 :\\[2mm]
    &\qquad\qquad u_t=\frac{\mathrm{e}^{-2t}(\mathrm{e}^{-2t}h-2u_xh')}{h''u_{xx}}+\mathrm{e}^{-t}F(\mathrm{e}^{2t}u_x),\\[2mm]
    & A_{3.6}^3\oplus A_1=\langle 1,u_t,tu_t+u,u_x \rangle :\quad u_t=u_x^2F(\frac{u_{xx}}{u_x}).
\end{align*}

$A_{3.7}\oplus A_1-$ invariant equations
\begin{align*}
    & A_{3.7}^1\oplus A_1=\langle 1,u_x,u-qxu_x,u_x^\frac1{1-q}h(t) \rangle,\ h'\neq0 :\quad u_t=-\frac{(q-1)^2h'u_x^2}{qhu_{xx}}+u_{x}^\frac1{1-q}F(t),\\[2mm]
    & A_{3.7}^1\oplus A_1=\langle 1,u_x,u-qxu_x,u_t \rangle :\quad u_t=u_{x}^\frac1{1-q}F(u_x^\frac{1-2q}{q-1}u_{xx}),\\[2mm]
    & A_{3.7}^2\oplus A_1=\langle 1,u_x,u_t+u-qxu_x,\mathrm{e}^{-t}h(\mathrm{e}^{(1-q)t}u_x) \rangle,\ h''\neq0,h(y)\neq y^\frac1{1-q} :\\[2mm]
    &\qquad\qquad u_t=\frac{(q-1)\mathrm{e}^{-qt}h'u_x+\mathrm{e}^{-t}h}{\mathrm{e}^{(1-2q)t}h''u_{xx}}+\mathrm{e}^{-t}F(\mathrm{e}^{(1-q)t}u_x),\\[2mm]
    & A_{3.7}^3\oplus A_1=\langle 1,u_t,-qtu_t+u,u_x \rangle :\quad u_t=u_x^{1-q}F(\frac{u_{xx}}{u_x}),\\[2mm]
    & A_{3.7}^4\oplus A_1=\langle u_t,1,-tu_t+qu,u_x \rangle :\quad
    u_t=u_x^\frac{q-1}qF(\frac{u_{xx}}{u_x}).
\end{align*}

$A_{3.8}\oplus A_1-$ invariant equations
\begin{align*}
    & A_{3.8}^1\oplus A_1=\langle 1,u_x,-x-uu_x,(1+u_x^2)^\frac12h(t) \rangle,\ h'\neq0:\\[2mm]
    & \qquad\qquad u_t=-\frac{h'(1+u_x^2)^2}{hu_{xx}}+(1+u_x^2)^\frac12F(t),\\[2mm]
    & A_{3.8}^1\oplus A_1=\langle 1,u_x,-x-uu_x,u_t \rangle :\quad u_t=(1+u_x^2)^\frac12F(\frac{u_{xx}}{(1+u_x^2)^\frac32}),\\[2mm]
    & A_{3.8}^2\oplus A_1=\langle 1,u_x,u_t-x-uu_x,(1+u_x^2)^\frac12h(t-\arctan{u_x}) \rangle,\ h'(h+h'')\neq0 :\\[2mm]
    & \qquad\qquad u_t=-\frac{h'(1+u_x^2)^2}{(h+h'')u_{xx}}+(1+u_x^2)^\frac12F(t-\arctan{u_x}).
\end{align*}

$A_{3.9}\oplus A_1-$ invariant equations
\begin{align*}
    & A_{3.9}^1\oplus A_1=\langle 1,u_x,(q-u_x)u-(1+qu_x)x,(1+u_x^2)^\frac12\mathrm{e}^{-q\arctan{u_x}}h(t) \rangle,\ h'\neq0 :\\[2mm]
    &\qquad\qquad u_t=-\frac{h'(1+u_x^2)^2}{(1+q^2)hu_{xx}}+(1+u_x^2)^\frac12\mathrm{e}^{-q\arctan{u_x}}F(t),\\[2mm]
    & A_{3.9}^1\oplus A_1=\langle 1,u_x,(q-u_x)u-(1+qu_x)x,u_t \rangle :\\[2mm]
    &\qquad\qquad u_t=(1+u_x^2)^\frac12\mathrm{e}^{-q\arctan{u_x}}F(\frac{\mathrm{e}^{-q\arctan{u_x}}u_{xx}}{(1+u_x^2)^\frac32}),\\[2mm]
    & A_{3.9}^2\oplus A_1=\langle 1,u_x,u_t+(q-u_x)u-(1+qu_x)x,\mathrm{e}^{-qt}(1+u_x^2)^\frac12h(t-\arctan{u_x}) \rangle :\\[2mm]
    &\qquad\qquad u_t=\frac{(qh-h')(1+u_x^2)^2}{(h+h'')u_{xx}}+\mathrm{e}^{-qt}(1+u_x^2)^\frac12F(t-\arctan{u_x}).
\end{align*}

\subsection{Four-dimensional non-decomposable algebras}

There exist ten non-isomorphic four-dimensional non-decomposable Lie
algebras, $A_{4.i}$ $(i=1,2,\cdots,10)$ (we give only non-zero commutation relations):
\begin{align*}
& A_{4.1}:\qquad [X_2,X_4]=X_1,\quad [X_3,X_4]=X_2;\\[2mm]
& A_{4.2}:\qquad [X_1,X_4]=qX_1,\quad [X_2,X_4]=X_2,\quad
[X_3,X_4]=X_2+X_3,\ q\neq0;\\[2mm]
& A_{4.3}:\qquad [X_1,X_4]=X_1,\quad [X_3,X_4]=X_2;\\[2mm]
& A_{4.4}:\qquad [X_1,X_4]=X_1,\quad [X_2,X_4]=X_1+X_2,\quad
[X_3,X_4]=X_2+X_3;\\[2mm]
& A_{4.5}:\qquad [X_1,X_4]=X_1,\quad [X_2,X_4]=qX_2,\quad
[X_3,X_4]=pX_3,\\[2mm]
& \qquad\qquad\ \ -1\leqslant p\leqslant q\leqslant1,\ pq\neq0;\\[2mm]
& A_{4.6}:\qquad [X_1,X_4]=qX_1,\quad [X_2,X_4]=pX_2-X_3,\quad
[X_3,X_4]=X_2+pX_3,\\[2mm]
& \qquad\qquad\ \ q\neq0,\ p\geqslant0;\\[2mm]
& A_{4.7}:\qquad [X_2,X_3]=X_1,\quad [X_1,X_4]=2X_1, \quad
[X_2,X_4]=X_2,\\[2mm]
&\qquad\qquad\ \ [X_3,X_4]=X_2+X_3;\\[2mm]
& A_{4.8}:\qquad [X_2,X_3]=X_1,\quad [X_1,X_4]=(1+q)X_1,\quad
[X_2,X_4]=X_2,\\[2mm]
& \qquad\qquad\ \ [X_3,X_4]=qX_3,\ |q|\leqslant1;\\[2mm]
& A_{4.9}:\qquad [X_2,X_3]=X_1,\quad [X_1,X_4]=2qX_1,\quad
[X_2,X_4]=qX_2-X_3,\\[3mm]
& \qquad\qquad\ \ [X_3,X_4]=X_2+qX_3,\ q\geqslant0;\\[2mm]
& A_{4.10}:\qquad [X_1,X_3]=X_1,\quad [X_2,X_3]=X_2,\quad
[X_1,X_4]=-X_2,\quad [X_2,X_4]=X_1.
\end{align*}
Each of  above algebras can be decomposed into a semi-direct sum
of a three-dimensional ideal $N$ and a one-dimensional Lie algebra.
Analysis of the commutation relations above shows that $N$ is of the
type $A_{3.1}$ for algebras $A_{4.i}\ (i=1,2,\cdots,6)$, of
type $A_{3.3}$ for algebras $A_{4.7},\ A_{4.8},\ A_{4.9}$, and
of type $A_{3.5}$ for algebra $A_{4.10}$. Thus we can extend
the already known realizations of three-dimensional Lie algebras to
obtain exhaustive description of the four-dimensional
non-decomposable solvable Lie algebras admitted by Eq.\eqref{eq}.
Below we present the final result, the lists of invariant equations
together with the corresponding symmetry algebras.

$A_{4.1}-$ invariant equations
\begin{align*}
    &A_{4.1}^1=\langle 1,u_x,u_t,-tu_x-x \rangle:\quad u_t=\frac{u_x^2}2+F(u_{xx}),\\[2mm]
    &A_{4.1}^2=\langle 1,u_x,\frac{u_x^2}2+h(t),-x \rangle,\ h'\neq0:\quad u_t=-\frac{h'}{u_{xx}}+F(t),\\[2mm]
    &A_{4.1}^3=\langle 1,(2u_x+h(t))^\frac12,u_x,-(2u_x+h(t))^\frac12x\rangle,\ h'\neq0:\quad u_t=\frac{(h+2u_x)h'}{2u_{xx}}+F(t).
\end{align*}

$A_{4.2}-$ invariant equations
\begin{align*}
    &A_{4.2}^1=\langle 1,u_x,u_t,-tu_t+qu-(t+x)u_x \rangle,\ q\neq1: \\[2mm]
    &\qquad\qquad u_t=\frac{u_x\ln{u_x}}{1-q}+u_xF(u_x^\frac{2-q}{q-1}u_{xx}),\\[2mm]
    &A_{4.2}^2=\langle 1,u_x,u_t,-tu_t+u-(t+x)u_x \rangle: \quad u_t=u_x\ln{u_{xx}}+F(u_x),\\[2mm]
    &A_{4.2}^3=\langle u_t,1,u_x,-qtu_t+u-(1+u_x)x \rangle :\quad u_t=\mathrm{e}^{(q-1)u_x}F(\mathrm{e}^{-u_x}u_{xx}),\\[2mm]
    &A_{4.2}^4=\langle 1,u_x,(\frac{\ln{u_x}}{1-q}+h(t))u_x,qu-xu_x \rangle,\ q\neq1 :\\[2mm]
    &\qquad\qquad u_t=\frac{(q-1)h'u_x^2}{u_{xx}}+u_x^\frac q{q-1}F(t),\\[2mm]
    &A_{4.2}^5=\langle \mathrm{e}^{(q-1)u_x}h(t),1,u_x,u-(1+u_x)x \rangle,\ q\neq1,h'\neq0 :\\[2mm]
    &\qquad\qquad u_t=-\frac{h'}{(q-1)^2hu_{xx}}+\mathrm{e}^{-u_x}F(t),\\[2mm]
    &A_{4.2}^6=\langle t^\frac12\mathrm{e}^{\lambda u_x},1,u_x,2(q-1-\lambda)tu_t-(1+u_x)x+u \rangle,\ \lambda\neq0,q-1 :\\[2mm]
    &\qquad\qquad u_t=-\frac{1}{2\lambda^2tu_{xx}}+t^\frac
    {\lambda^2-2\lambda-q^2+2q-2}{4(q-1-\lambda)^2}F(2u_x+\frac{\ln{t}}{1-q+\lambda}).
\end{align*}

$A_{4.3}-$ invariant equations
\begin{align*}
    &A_{4.3}^1=\langle 1,u_x,u_t,u-tu_x \rangle:\quad u_t=-u_x\ln{u_x}+u_xF(\frac{u_{xx}}{u_x}),\\[2mm]
    &A_{4.3}^2=\langle u_t,1,u_x,-tu_t-x \rangle:\quad u_t=\mathrm{e}^{u_x}F(u_{xx}),\\[2mm]
    &A_{4.3}^3=\langle 1,u_x,(-\ln{u_x}+h(t))u_x,u \rangle,\ h'\neq0:\quad u_t=\frac{h'u_x^2}{u_{xx}}+u_xF(t),\\[2mm]
    &A_{4.3}^4=\langle \mathrm{e}^{u_x}h(t),1,u_x,-x \rangle,\ h'\neq0:\quad u_t=-\frac{h'}{hu_{xx}}+F(t),\\[2mm]
    &A_{4.3}^5=\langle t^\frac12\mathrm{e}^{\lambda u_x},1,u_x,2(1-\lambda)tu_t-x \rangle,\ \lambda\neq0,1 :\\[2mm]
    &\qquad\qquad u_t=-\frac1{2\lambda^2tu_{xx}}+\frac1tF(2u_x+\frac{\ln{t}}{\lambda-1}).
\end{align*}

$A_{4.4}-$ invariant equations
\begin{align*}
    &A_{4.4}^1=\langle 1,u_x,u_t,-tu_t-(t+x)u_x+u-x \rangle:\quad u_t=\frac{u_x^2}2+F(\mathrm{e}^{-u_x}u_{xx}),\\[2mm]
    &A_{4.4}^2=\langle 1,u_x,\frac{u_x^2}2+h(t),u-(1+u_x)x \rangle,\ h'\neq0 :\quad u_t=-\frac{h'}{u_{xx}}+\mathrm{e}^{-u_x}F(t),\\[2mm]
    &A_{4.4}^3=\langle 1,(2u_x+h(t))^\frac12,u_x,u-((2u_x+h(t))^\frac12+u_x)x \rangle,\ h'\neq0:\\[2mm]
    &\qquad\qquad
    u_t=\frac{h'(2u_x+h)}{2u_{xx}}+\mathrm{e}^{-(2u_x+h(t))^\frac12}F(t).
\end{align*}

$A_{4.5}-$ invariant equations
\begin{align*}
    &A_{4.5}^1=\langle 1,u_x,u_t,-ptu_t+u-qxu_x \rangle,\ q\neq1:\quad u_t=u_x^\frac{p-1}{q-1}F(u_x^\frac{1-2q}{q-1}u_{xx}),\\[2mm]
    &A_{4.5}^2=\langle 1,u_x,u_t,-ptu_t+u-xu_x \rangle:\quad u_t=u_{xx}^{p-1}F(u_x),\\[2mm]
    &A_{4.5}^3=\langle 1,u_t,u_x,-qtu_t+u-pxu_x \rangle,\ p\neq1:\quad u_t=u_x^\frac{q-1}{p-1}F(u_x^\frac{1-2p}{p-1}u_{xx}),\\[2mm]
    &A_{4.5}^4=\langle u_t,1,u_x,-tu_t+qu-pxu_x \rangle,\ p\neq q:\quad u_t=u_x^\frac{1-q}{p-q}F(u_x^\frac{q-2p}{p-q}u_{xx}),\\[2mm]
    &A_{4.5}^5=\langle u_t,1,u_x,-tu_t+qu-qxu_x \rangle,\ q\neq 1:\quad u_t=u_{xx}^\frac{1-q}qF(u_x),\\[2mm]
    &A_{4.5}^6=\langle 1,u_x,h(t,u_x),u-xu_x \rangle,\ h_th_{u_xu_x}\neq0:\quad u_t=-\frac{h_t}{h_{u_xu_x}u_{xx}},\\[2mm]
    &A_{4.5}^7=\langle 1,u_x,u_x^\frac{p-1}{q-1}h(t),u-qxu_x \rangle,\ -1\le p<q<1,h'\neq0:\\[2mm]
    &\qquad\qquad u_t=\frac{(q-1)^2u_x^2h'}{(1-p)(p-q)u_{xx}h}+u_x^\frac1{1-q}F(t),\\[2mm]
    &A_{4.5}^8=\langle 1,u_x^\frac{q-1}{p-1}h(t),u_x,u-pxu_x) \rangle,\ -1\le p<q<1,h'\neq0:\\[2mm]
    &\qquad\qquad u_t=\frac{(p-1)^2u_x^2h'}{(q-1)(p-q)u_{xx}h}+u_x^\frac1{1-p}F(t),\\[2mm]
    &A_{4.5}^9=\langle u_x^\frac{1-q}{p-q}h(t),1,u_x,qu-pxu_x \rangle,\ -1\le p<q<1,h'\neq0:\\[2mm]
    &\qquad\qquad u_t=\frac{(p-q)^2u_x^2h'}{(p-1)(1-q)u_{xx}h}+u_x^\frac
    q{q-p}F(t).
            \end{align*}

$A_{4.6}-$ invariant equations
\begin{align*}
    &A_{4.6}^1=\langle u_t,1,u_x,-qtu_t+(p-u_x)u-(1+pu_x)x \rangle:\\[2mm]
    &\qquad\qquad u_t=(1+u_x^2)^\frac12\mathrm{e}^{(q-p)\arctan{u_x}}F(\frac{\mathrm{e}^{-p\arctan{u_x}}u_{xx}}{(1+u_x^2)^\frac32}),\\[2mm]
    &A_{4.6}^2=\langle (1+u_x^2)^\frac12\mathrm{e}^{(q-p)\arctan{u_x}}h(t),1,u_x,(p-u_x)u-(1+pu_x)x \rangle:\\[2mm]
    &\qquad\qquad u_t=-\frac{(1+u_x^2)^2h'}{(p^2+q^2-2pq+1)hu_{xx}}+(1+u_x^2)^\frac12\mathrm{e}^{-p\arctan{u_x}}F(t).
\end{align*}

$A_{4.7}-$ invariant equations
\begin{align*}
    & A_{4.7}^1=\langle 1,u_x,u_t-x,(-t+\lambda)u_t+\frac{t^2}2-(t+x)u_x+2u \rangle :\\[2mm]
    &\qquad\qquad u_t=\omega F(u_{xx})+\omega \ln{\omega}-\lambda,\quad \omega=t-u_x-\lambda,\\[2mm]
    & A_{4.7}^2=\langle 1,u_x,-x,2u-xu_x-\frac{u_x^2}2 \rangle :\quad u_t=\mathrm{e}^\frac2{u_{xx}}F(t),\\[2mm]
    & A_{4.7}^3=\langle 1,u_x,-x,u_t+2u-xu_x-\frac{u_x^2}2 \rangle :\quad u_t=\mathrm{e}^{-2t}F(t+\frac1{u_{xx}}),\\[2mm]
    & A_{4.7}^4=\langle 1,u_x-t,-u_t,-(t+\lambda)u_t+2u+(t-x)u_x+\lambda x-\frac{t^2}2 \rangle :\\[2mm]
    &\qquad\qquad u_t=(u_x+\lambda)\ln{(u_x+\lambda)}+(u_x+\lambda)F(u_{xx})+x+\lambda.
\end{align*}

$A_{4.8}-$ invariant equations
\begin{align*}
    & A_{4.8}^1=\langle 1,u_x,u_t-x,(-qt+\lambda)u_t+(1+q)u-xu_x \rangle,\ q\neq0 :\\[2mm]
    &\qquad\qquad u_t=\omega^\frac1qF(\omega^\frac{1-q}qu_{xx}),\quad \omega=q(t-u_x)-\lambda,\\[2mm]
    & A_{4.8}^2=\langle 1,u_x,u_t-x,\lambda u_t+u-xu_x \rangle,\ \lambda\neq0 :\\[2mm]
    &\qquad\qquad u_t=\omega F(\omega u_{xx}),\quad \omega=\mathrm{e}^\frac{u_x-t}\lambda, \\[2mm]
    & A_{4.8}^3=\langle 1,u_x,u_t-x,u-xu_x \rangle :\quad u_t=\frac{F(u_x-t)}{u_{xx}},\\[2mm]
    & A_{4.8}^4=\langle 1,u_t-x,-u_x,(-t+\lambda)u_t+(1+q)u-qxu_x \rangle :\\[2mm]
    &\qquad\qquad u_t=\omega^qF(\omega^{q-1}u_{xx}),\quad \omega=t-u_x-\lambda,\\[2mm]
    & A_{4.8}^5=\langle 1,u_x,-x,(1+q)u-xu_x \rangle,\ |q|<1:\quad u_t=u_{xx}^\frac{q+1}{q-1}F(t),\\[2mm]
    & A_{4.8}^6=\langle 1,u_x,-x,u_t+(1+q)u-xu_x \rangle:\quad u_t=\mathrm{e}^{-(1+q)t}F(\mathrm{e}^{(q-1)t}u_{xx}),\\[2mm]
    & A_{4.8}^7=\langle 1,u_t,u_x-t,-(t+\lambda)u_t+(1+q)u+(\lambda-qu_x)x \rangle :\\[2mm]
    &\qquad\qquad u_t=x+(u_x+\lambda)^qF((u_x+\lambda)^{q-1}u_{xx}),\\[2mm]
    & A_{4.8}^8=\langle 1,u_x-t,-u_t,-(qt+\lambda)u_t+(1+q)u+(\lambda-u_x)x \rangle,\ q\neq0 :\\[2mm]
    &\qquad\qquad u_t=x+(qu_x+\lambda)^\frac1qF((qu_x+\lambda)^\frac{1-q}qu_{xx}),\\[2mm]
    & A_{4.8}^9=\langle 1,u_x-t,-u_t,-\lambda u_t+u+(\lambda-u_x)x \rangle,\ q\lambda\neq0:\\[2mm]
    &\qquad\qquad u_t=x+\mathrm{e}^\frac{u_x}\lambda F(\mathrm{e}^\frac{u_x}\lambda u_{xx}),\\[2mm]
    & A_{4.8}^{10}=\langle 1,u_x-t,-u_t,u-u_xx \rangle :\quad
    u_t=x+\frac{F(u_x)}{u_{xx}}.
\end{align*}

$A_{4.9}-$ invariant equations
\begin{align*}
    &A_{4.9}^1=\langle 1,u_x,-x,q(2u-xu_x)-\frac{x^2+u_x^2}2 \rangle:\quad u_t=\mathrm{e}^{-2q\arctan{u_{xx}}}F(t),\\[2mm]
    &A_{4.9}^2=\langle 1,u_x,-x,u_t+q(2u-xu_x)-\frac{x^2+u_x^2}2 \rangle:\\[2mm]
    &\qquad\qquad u_t=\mathrm{e}^{-2qt}F(t-\arctan{u_{xx}}).
\end{align*}

$A_{4.10}-$ invariant equations
\begin{align*}
    &A_{4.10}^1=\langle 1,u_x,u_t-xu_x+u,\lambda u_t-uu_x-x \rangle:\\[2mm]
    &\qquad\qquad u_t=(1+u_x^2)^\frac12\mathrm{e}^{-t+\lambda\arctan{u_x}}F(\frac{\mathrm{e}^{-t+\lambda\arctan{u_x}}u_{xx}}{(1+u_x^2)^\frac32}),\\[2mm]
    &A_{4.10}^2=\langle 1,u_x,u-xu_x,-uu_x-x \rangle:\quad u_t=\frac{(1+u_x^2)^2}{u_{xx}}F(t),\\[2mm]
    &A_{4.10}^3=\langle 1,u_x,u-xu_x,u_t-uu_x-x \rangle:\quad
    u_t=\frac{(1+u_x^2)^2}{u_{xx}}F(t-\arctan{u_x}).
\end{align*}

\section{Classification of equations invariant under
the algebras having nontrivial Levi factor}

In this section, utilizing the classification results of Eq. \eqref{eq} admitting semi-simple symmetry algebras, we describe \eqref{eq} which are invariant with respect  to Lie algebras having Levi decomposition. These algebras split into the flowing two non-isomorphic algebras

$\bullet$ Lie algebras which are decomposable into direct sums of semi-simple and solvable Lie algebras,

$\bullet$ Lie algebras which are semi-direct sums of a Levi factor and nonzero radical.

Now we consider these two categories separately.

\subsection{Direct sums of semi-simple and solvable Lie algebras}

Here we utilize the results of classification of inequivalent equations \eqref{eq} admitting semi-simple
symmetry algebras to describe PDEs \eqref{eq} whose symmetry algebras are decomposable into direct sum of
semi-simple and solvable Lie algebras.

Consider the case of the ${\frak{sl}}^1(2,\mathbf{R})$ invariant equation. We look for possible extensions
of the realization ${\frak{sl}}^1(2,\mathbf{R})$ by function \eqref{g} which commute with its basis functions.
Analysis of the commutation conditions yields the general generating function
\begin{equation}\label{sl1}
  g=\alpha(t)u_t+\phi(t)u_x,
\end{equation}
where $\phi$ is an arbitrary real-valued function. Now we need to construct all possible solvable Lie algebras with
the generating function \eqref{sl1}. Here we skip intermediate computation and only list the final results. Solvable Lie algebras
realized by \eqref{sl1} are isomorphic to two one-dimensional algebras $\langle u_t \rangle$, $\langle \phi(t)u_x \rangle$ with $\dot{\phi}\neq0$ and one two-dimensional algebras $\langle u_t,-tu_t+\lambda u_x \rangle$ of type $A_{2.2}$.

Substituting above generating functions into the classifying equation \eqref{ce} and solving the obtained PDEs,
yield the corresponding invariant equation
\begin{align*}
    &{\frak{sl}}^1(2,\mathbf{R})\oplus\langle u_t \rangle:\\[2mm]
    &\qquad\qquad u_t=u_xF(-x+\mathrm{arctanh}{\frac{u_{xx}}{u_x}}),\\[2mm]
    &{\frak{sl}}^1(2,\mathbf{R})\oplus\langle \phi(t)u_x \rangle,\ \dot{\phi}\neq0:\\[2mm]
    &\qquad\qquad u_t=-\frac{u_x}{2\phi}(\dot{\phi}\ln{\frac{u_{xx}-u_x}{u_{xx}+u_x}}-2\dot{\phi}x)+u_xF(t),\\[2mm]
    &{\frak{sl}}^1(2,\mathbf{R})\oplus\langle u_t,-tu_t+\lambda u_x \rangle,\ \lambda\neq0:\\[2mm]
    &\qquad\qquad u_t=Cu_x\mathrm{e}^{\frac x\lambda}(\frac{u_{xx}-u_x}{u_{xx}+u_x})^\frac1{2\lambda}.
\end{align*}

Analogously, we can construct invariant equations \eqref{eq} admitting algebras which are direct sum of ${\frak{sl}}^2(2,\mathbf{R})$
and solvable algebras,
\begin{align*}
    &{\frak{sl}}^2(2,\mathbf{R})\oplus\langle u_t \rangle:\\[2mm]
    &\qquad\qquad u_t=u_xF(x+\arctan{\frac{u_{xx}}{u_x}}),\\[2mm]
    &{\frak{sl}}^2(2,\mathbf{R})\oplus\langle \phi(t)u_x \rangle,\ \dot{\phi}\neq0:\\[2mm]
    &\qquad\qquad u_t=-\frac{u_x\dot{\phi}}{\phi}(x+\arctan{\frac{u_{xx}}{u_x}})+u_xF(t),\\[2mm]
    &{\frak{sl}}^2(2,\mathbf{R})\oplus\langle u_t,-tu_t+\lambda u_x \rangle,\ \lambda\neq0:\\[2mm]
    &\qquad\qquad u_t=Cu_x\mathrm{e}^{\frac{x+\arctan{\frac{u_{xx}}{u_x}}}\lambda}.
\end{align*}

For ${\frak{sl}}^3(2,\mathbf{R})$, there exist two solvable algebras $\langle 1\rangle$ and $\langle 1,u\rangle$
which can commute with it. Below we list the corresponding invariant equations.
\begin{align*}
    &{\frak{sl}}^3(2,\mathbf{R})\oplus\langle 1\rangle:\\[2mm]
    &\qquad\qquad u_t=-x^2u_x+\frac1{u_x}F(\frac{u_{xx}}{u_x^2}).,\\[2mm]
    &{\frak{sl}}^3(2,\mathbf{R})\oplus\langle 1,u\rangle:\\[2mm]
    &\qquad\qquad u_t=-x^2u_x+C\frac{u_x^3}{u_{xx}^2}.
\end{align*}

A similar analysis of extensions of the realization of algebra $\frak{so}(3)$ yields one more invariant equations,
\begin{align*}
    &{\frak{so}}^1(3)\oplus\langle u_t\rangle:\\[2mm]
    &\qquad\qquad u_t=(\sec^2{x}+u_x^2)^\frac12 F(\frac{u_{xx}\cos{x}-(2+u_x^2\cos^2{x})u_x\sin{x}}{(1+u_x^2\cos^2{x})^\frac32}).
\end{align*}
Under arbitrary $F$ and $C$, the given algebras are maximal in Lie's sense invariance algebras of the corresponding
equations.

\subsection{Semi-direct sums of semi-simple and solvable Lie algebras}

To perform classification of Eq. \eqref{eq} whose invariance algebras are isomorphic to semi-direct sum of semi-simple and solvable Lie algebras, we need to apply a two-step approach, following Ref. \cite{zhd07}. Firstly, using the classification results of lower dimensional Lie algebras which are semi-direct sum of Levi factor and solvable radical \cite{tur88}, we describe all invariant equations containing arbitrary functions of two, one arguments or arbitrary constants. According to Ref. \cite{tur88}, without loss of generality, here we can restrict our consideration to Lie algebras having Levi decomposition $\frak{sl}(2,\mathbf{R})\sdsum A_{2.1}$ and $\frak{so}(3)\sdsum A_{3.1}$. Secondly, for equations having arbitrary functions of one variable or arbitrary constants, we apply Ovsiannikov method \cite{ovs82} to finalize the classification.

Now we consider algebra $\frak{sl}(2,\mathbf{R})\sdsum A_{2.1}$ first. Let $\frak{sl}(2,\mathbf{R})=\langle g_1,g_2,g_3 \rangle$ and $A_{2.1}=\langle g_4,g_5 \rangle$, the nonzero commutation relations of this algebra read
\begin{equation*}
\begin{array}{c}
  [g_1,g_2]=g_1,\quad [g_1,g_3]=2g_2,\quad [g_2,g_3]=g_3,\quad [g_1,g_5]=g_4,\\[2mm]
  [g_2,g_4]=-\frac12g_4,\quad [g_2,g_5]=\frac12 g_5,\quad [g_3,g_4]=-g_5.
\end{array}
\end{equation*}
Here we provide full calculation details for algebra ${\frak{sl}}^1(2,\mathbf{R})\sdsum A_{2.1}$, other cases are handled in the same way. Inserting $g_1=1,\ g_2=u,\ g_3=u^2-u_x^2$ and $g_4,g_5$ of the general form \eqref{g} into the commutation relations of $\frak{sl}(2,\mathbf{R})\sdsum A_{2.1}$ and solving resulted equations, we obtain
\begin{equation*}
    g_4=\phi(t)(\mathrm{e}^{-x}u_x)^{\frac12},\quad g_5=\phi(t)(\mathrm{e}^{-x}u_x)^{\frac12}(u+u_x),
\end{equation*}
where $\phi(t)$ is an arbitrary nonzero function. Applying contact transformation
\begin{equation*}
    \tilde{t}=T(t),\quad \tilde{x}=x+Y(t),\quad \tilde{u}=u,
\end{equation*}
which preserves ${\frak{sl}}^1(2,\mathbf{R})$ to characteristic functions obtained above yields
\begin{equation*}
  \tilde{g}_4 \rightarrow g_4=\frac{\alpha(T)}{\mathrm{e}^\frac{Y(t)}2}(\mathrm{e}^{-x}u_x)^{\frac12},\quad
  \tilde{g}_5 \rightarrow g_5=\frac{\alpha(T)}{\mathrm{e}^\frac{Y(t)}2}(\mathrm{e}^{-x}u_x)^{\frac12}(u+u_x).
\end{equation*}
Choosing $T=t$ and $Y(t)$ as the solution of equation $|\alpha(t)|=\mathrm{e}^\frac{Y(t)}2$, we arrive at the only inequivalent extension of algebra ${\frak{sl}}^1(2,\mathbf{R})$, ie.
\begin{equation*}
    {\frak{sl}}^1(2,\mathbf{R}) \sdsum \langle (\mathrm{e}^{-x}u_x)^{\frac12},(\mathrm{e}^{-x}u_x)^{\frac12}(u+u_x) \rangle,
\end{equation*}
the corresponding invariant equation is
\begin{equation}\label{sleq}
    u_t=\frac{u_x(u_{xx}+u_x)^\frac13\mathrm{e}^{-\frac23x}}{(u_{xx}-u_x)^\frac13}F(t).
\end{equation}

However, this five-dimensional Lie algebra is not maximal. To find the most extensive symmetry algebra, we normalize Eq. \eqref{sleq} firstly. In view of $F(t)\neq0$, we make the change of variables,
\begin{equation*}
    \tilde{t}=\int F(t)\mathrm{d}t,\quad \tilde{x}=x,\quad\tilde{u}=u,
\end{equation*}
to Eq. \eqref{sleq} and get the equation
\begin{equation*}
    u_t=\frac{u_x(u_{xx}+u_x)^\frac13\mathrm{e}^{-\frac23x}}{(u_{xx}-u_x)^\frac13}.
\end{equation*}
Note that here we drop the bars. Now applying Lie infinitesimal algorithm directly, we obtain the maximal invariance algebra of Eq. \eqref{sleq} with $F=1$. It is the seven-dimensional Lie algebra
\begin{equation*}
    {\frak{sl}}^1(2,\mathbf{R}) \sdsum \langle (\mathrm{e}^{-x}u_x)^{\frac12},(\mathrm{e}^{-x}u_x)^{\frac12}(u+u_x), u_t,-\frac43tu_t-2u_x\rangle,
\end{equation*}
which is isomorphic to $\frak{sl}(2,\mathbf{R}) \sdsum A_{4.5}$ with $q=1$ and $p=4/3$.

Analysis of algebra ${\frak{sl}}^2(2,\mathbf{R})$ shows that it can not be extended up to an invariance algebra of \eqref{eq} isomorphic to $\frak{sl}(2,\mathbf{R}) \sdsum A_{2.1}$. Algebra ${\frak{sl}}^3(2,\mathbf{R})$ admits a extension to algebra $\frak{sl}(2,\mathbf{R}) \sdsum A_{2.1}$, the algebra and its invariant equation are given as
\begin{equation*}
\begin{array}{c}
    {\frak{sl}}^3(2,\mathbf{R}) \sdsum\langle\frac{xu_x+u}{\sqrt{u_x}},-\frac{(tx+1)u_x+tu}{\sqrt{u_x}}\rangle:\\[2mm]
    u_t=-2\lambda\frac{u_x}{u_{xx}}+\frac{u^2+\lambda u}{u_x}-x^2u_x,\quad\lambda\neq0.
\end{array}
\end{equation*}
Note that this five-dimensional Lie algebra presented above is maximal in Lie's sense.

For the case of algebra $\frak{so}(3)\sdsum A_{3.1}$, we let $\frak{so}(3)=\langle g_1,g_2,g_3 \rangle$ and $A_{3.1}=\langle g_4,g_5,g_6 \rangle$, the nonzero commutation relations between $\frak{so}(3)$ and $A_{3.1}$ are
\begin{equation*}
\begin{array}{c}
  [g_1,g_5]=g_6,\quad [g_1,g_6]=-g_5,\quad [g_2,g_4]=-g_6,\\[2mm]
  [g_2,g_6]=g_4,\quad [g_3,g_4]=g_5,\quad [g_3,g_5]=-g_4.
\end{array}
\end{equation*}
Now we choose
\begin{equation*}
    g_1=1,\quad g_2=\tan{x}\sin{u}-u_x\cos{u},\quad g_3=\tan{x}\cos{u}+u_x\sin{u},
\end{equation*}
and $g_4,g_5,g_6$ of the form \eqref{g}. Substituting them into the commutation relations of algebra $\frak{so}(3)\sdsum A_{3.1}$, we easily get that
\begin{equation*}
    g_4=g_5=g_6=0.
\end{equation*}
Thus Eq. \eqref{eq} can not admit a symmetry algebra having Levi decomposition $\frak{so}(3)\sdsum A_{3.1}$.

\section{Concluding remarks}
In this paper, we develop an algebraic approach for group classification of contact symmetries and perform contact symmetry classification of the second-order evolution equation \eqref{eq}.
As a result, the broad classes of invariant equations \eqref{eq} are constructed
together with their maximal contact symmetry algebras. Symmetry properties of these equations can be
briefly summarized as follows. We obtain

$\bullet$ two inequivalent equations admitting one-dimensional Lie algebra.

$\bullet$ four equations which admit semi-simple Lie algebras.

$\bullet$ five equations admitting two-dimensional Lie algebras.

$\bullet$ twenty-six equations admitting three-dimensional solvable Lie algebras.

$\bullet$ eighty-eight equations admitting four-dimensional solvable Lie algebras.

$\bullet$ eleven equations admitting admitting symmetry algebras having nontrivial Levi factor.

\end{document}